\documentclass[a4paper,fleqn,usenatbib]{mnras}
\usepackage[T1]{fontenc}
\usepackage{ae,aecompl}
\usepackage{graphicx}	
\usepackage{amsmath}	
\usepackage{amssymb}	
\usepackage{epstopdf}
\usepackage{hyperref}

\newcommand{\grs}{{GRS\,1739--278}}
\newcommand{\swiftx}{{\it Swift}/{\rm XRT}}
\newcommand{\swiftb}{{\it Swift}/{\rm BAT}}
\newcommand{\xmm}{{\it XMM-Newton}}
\newcommand{\nustar}{{\it NuSTAR}}
\newcommand{\integral}{{\it INTEGRAL}}
\newcommand{\maxi}{{\rm MAXI}}

\newcommand{\ferg}{{erg~cm$^{-2}$~s$^{-1}$}}

\newcommand{\quality}{Q}

\title[Timing properties of \grs]{Studying temporal variability of \grs\ during the 2014 outburst}

\author[I. A. Mereminskiy et al.]
{Ilya A. Mereminskiy,$^{1}$\thanks{E-mail: i.a.mereminskiy@gmail.com}
Andrey N. Semena,$^{1}$
Sergey D. Bykov,$^{1,2}$ \newauthor
Ekaterina V. Filippova,$^{1}$
Alexander A. Lutovinov$^{1,3}$  and 
Juri Poutanen$^{1,4,5}$
\\
$^{1}$Space Research Institute, Russian Academy of Sciences, Profsoyuznaya 84/32, 117997 Moscow, Russia\\
$^{2}$Bauman Moscow State Technical University, Baumanskaya 2, 105005 Moscow, Russia\\
$^{3}$Higher School of Economics, Myasnitskaya 20, 101000 Moscow, Russia\\
$^{4}$Tuorla observatory, Department of Physics and Astronomy, FI-20014 University of Turku, Finland\\
$^{5}$Nordita, KTH Royal Institute of Technology and Stockholm University, Roslagstullsbacken 23, SE-10691 Stockholm, Sweden\\
}

\date{Accepted XXX. Received YYY; in original form ZZZ}

\pubyear{2018}

\begin{document}
\label{firstpage}
\pagerange{\pageref{firstpage}--\pageref{lastpage}}
\maketitle

\begin{abstract}
We report a discovery of low-frequency quasi periodic oscillation at 0.3--0.7 Hz in the power spectra of the accreting black hole \grs\ in the hard-intermediate state during its 2014 outburst based on the \nustar\ and \swiftx\ data. 
The QPO frequency strongly evolved with the source flux during the \nustar\ observation.
The source spectrum became softer with rising QPO frequency and simultaneous increasing of the power-law index and decreasing of the cut-off energy.
In the power spectrum, a prominent harmonic is clearly seen together with the main QPO peak. 
The fluxes in the soft and the hard X-ray bands are coherent, however, the coherence drops for the energy bands separated by larger gaps.
The phase-lags are generally positive (hard) in the 0.1--3~Hz frequency range, and negative below 0.1~Hz. The accretion disc inner radius estimated with the relativistic reflection spectral model appears to be $R_{\rm in} < 7.3 R_{\rm g}$.
In the framework of the relativistic precession model, in order to satisfy the constraints from the observed QPO frequency and the accretion disc truncation radius, a massive black hole with $M_{\rm BH} \approx 100$M$_\odot$ is required.
\end{abstract}


\section{Introduction}
\label{sec:intro} 

A study of X-ray variability in accreting compact objects provides a broad view on processes that take place in the vicinity of these objects.
This is true for both long (days and weeks) time-scales, when one speaks about state changes through outbursts of the transients sources \citep[see e.g.][]{homan05, heil15}, and short time-scales (down to milliseconds), when the subject under consideration is quasi-periodic oscillations (QPOs) and the broad-band stochastic noise. 
A simultaneous usage of spectral and timing data can help to better constrain the geometry of accretion flows around compact objects and infer the processes, which are responsible for generation of the observed spectral and timing characteristics.

Many aspects of the spectral and timing evolution of black hole (BH) X-ray transients during outbursts can be explained in the framework of the truncated disc model \citep[see e.g.][for reviews]{ZG04,DGK07,2014SSRv..183...61P}.
In this model, the low/hard state (LHS) is characterized by the geometrically thin cold standard \citep{shakura73} disc truncated at some large radius and the inner parts of the accretion flow  occupied by the  hot thick two-temperature disc  \citep{1975ApJ...199L.153E, 1976ApJ...204..187S, 1995ApJ...452..710N}. 
The X-ray spectra are dominated by thermal Comptonization of soft seed photons either from the cold disc or from the synchrotron  radiation of non-thermal electrons \citep{PV09,MB09,VPV13,PVZ18}.
Decreasing of the truncation radius leads to the transition to the soft state \citep{Esin97,PKR97}. 
In the high/soft state (HSS), the X-ray spectra are dominated by the emission from the standard disc extending down to the innermost stable circular orbit (ISCO), while the high-energy emission is produced by Comptonization of the disc photons in the non-thermal corona \citep{1998PhST...77...57P,gierlinski99,ZGP01} as supported by the frequency-resolved spectroscopy \citet{2001MNRAS.321..759C}. 
The truncation disc model finds support not only from the X-ray spectral data, but also \citep[see review in][]{2014SSRv..183...61P} from the X-ray timing properties such as time lags \citep{2001MNRAS.327..799K}, the Fourier frequency-resolved spectra \citep{RGC99,GCR99,2000MNRAS.316..923G} as well as from the correlated optical-X-ray variability \citep{VPV11,VPI13} and the phase-connected QPOs \citep{VRD15}.  

\citet{1997MNRAS.292..679L} proposed that the observed strong variability (seen as the broad-band noise in the power spectra) is produced by the stochastic variations of the viscosity. 
In this propagating fluctuation model, the broad-band noise is a product of the noise signals from different radii of the accretion flow, with their own time-scales \citep[see, also,][]{2006MNRAS.367..801A, 2013MNRAS.434.1476I}. 
Therefore, the shape of the broad-band noise is determined by the physical and geometrical properties of the accretion flow. 
In particular, it was argued that the power spectra break frequency is connected with the inner edge of the accretion flow. 

Another feature, frequently observed in the power spectra of X-ray binaries, is the QPOs, manifesting themselves as narrow Lorentzian components. 
The low-frequency (LF) QPOs are better studied, because they occur at moderate frequencies of 0.1--10~Hz. 
These QPOs are ubiquitous: they were found in systems with neutron stars and BHs \citep{wijnands99}, cataclysmic variables \citep{mauche02} and active galactic nuclei \citep{gierlinski08}. 
Few types of LF QPOs are distinguished, based on the shape of the power spectra \citep[see e.g.][]{casella05}, but their origin remains still unclear.

The so-called type-C LF QPOs are typically found in X-ray BH transients during the initial rising part of the outbursts and the transition to the HSS, i.e. in the LHS and in the hard intermediate state (HIMS) \citep{tanaka96,grebenev97, remillard06, belloni10}; sometimes they are seen at higher frequencies ($\approx$30~Hz) after a transition to the HSS. 
These QPOs are easy to detect and study, because they are prominent, with the relative rms of $\approx10$ per cent  \citep{casella05}. 
These QPOs were proposed to result from the Lense-Thirring precession of inner parts of the accretion disc \citep{FB07,ingram09,VPI13}, oscillations of a standing shock \citep{molteni96} and the accretion rate modulation caused by different phenomena \citep{tagger99,cabanac10}. 
In some models, particularly in the Lense-Thirring precession model, the observed frequency strongly depends on the truncation radius of the cold disc, at which it transforms into a geometrically thick, optically thin hot flow. 

Recent advances in simulations of the reflected emission \citep{garcia14}, arising due to the scattering and absorption of the hard X-ray photons in the cold accretion disc, led to a possibility to study the geometry of the accretion flow.  
For such a study  it is essential to obtain a broad-band X-ray spectrum with the high energy resolution as the reflected emission manifests itself by the presence of a prominent, wide and asymmetric iron $K_{\alpha}$ fluorescent emission line at $\sim$6.4 keV and the Compton-hump at 20--30 keV. 
Combination of the reflection models with the self-consistent modelling of the spectral formation allow us then to put constraints on the geometry of the X-ray emitting region \citep{PVZ18}. 
The X-ray timing analysis provides additional information on the location of the component responsible for the variability and on the distance scale.
This task presents a challenge, that can be solved only by the instruments that provide the means to measure the  broad-band spectrum with good resolution as well as the  timing characteristics. 
\nustar\ \citep{harrison13_nust}, launched in 2013, is currently the best available instrument for such studies. 
\xmm\ and {\it NICER} can also be used, although their energy range reaching only up to $\sim$12~keV limits their capabilities to measure hard tails and contribution of the Compton-hump.

In this paper we report the discovery of the type-C QPOs in the HIMS of the Galactic BH \grs\ and present a detailed study of its X-ray variability, along with the spectral evolution using the \nustar\ and \swiftx\ data.
In Section~\ref{sec:source} we describe the source under investigation and in Section~\ref{sec:data} we describe its 2014 outburst, the data analysis and the reduction procedure. 
The properties of the energy spectrum of \grs\  and its evolution with time are presented in Section~\ref{sec:spectra}.
In Section~\ref{sec:timing} we investigate the timing properties of the source and evaluate the obtained results in the framework of the propagating fluctuations model.
In Section~\ref{sec:disc} we discuss the obtained results and extract some physical quantities using the Lense-Thirring precession model for the QPOs.

\section{GRS 1739--278}
\label{sec:source} 

\grs\ is a typical X-ray nova, discovered during an outburst in 1996 \citep{paul96} by the SIGMA \citep{paul91} telescope onboard the {\it GRANAT} space observatory \citep[see e.g.][]{grebenev93}.
Using {\it ROSAT} measurement of the absorption column,  \citet{greiner96} inferred the distance to the source of 6--8.5~kpc, indicating that the source may belong to the Galactic bulge. 
It should be noted that \citet{greiner96} used the X-ray halo size to assess the obscuration column density and the value of the mean extinction per parsec from \citet{1973asqu.book.....A} to estimate the distance. 
Although this $N_{\rm H}$ estimation appears to be quite precise, it is larger than the new measurements of the line-of-sight absorption in the Galaxy towards the source \citep{1990ARA&A..28..215D, 2005A&A...440..775K, 2006A&A...453..635M, 2014A&A...566A.120S}. 
It means that the source has either some intrinsic obscuration or additional line-of-sight obscuration,  and its distance cannot be easily constrained from the measurements of $N_{\rm H}$.
Nevertheless, in this work we will assume  the distance to \grs\ of 8~kpc, given that the source is projected on to the Galactic Bulge.  

\begin{figure*}
\centerline{\includegraphics[scale=0.75]{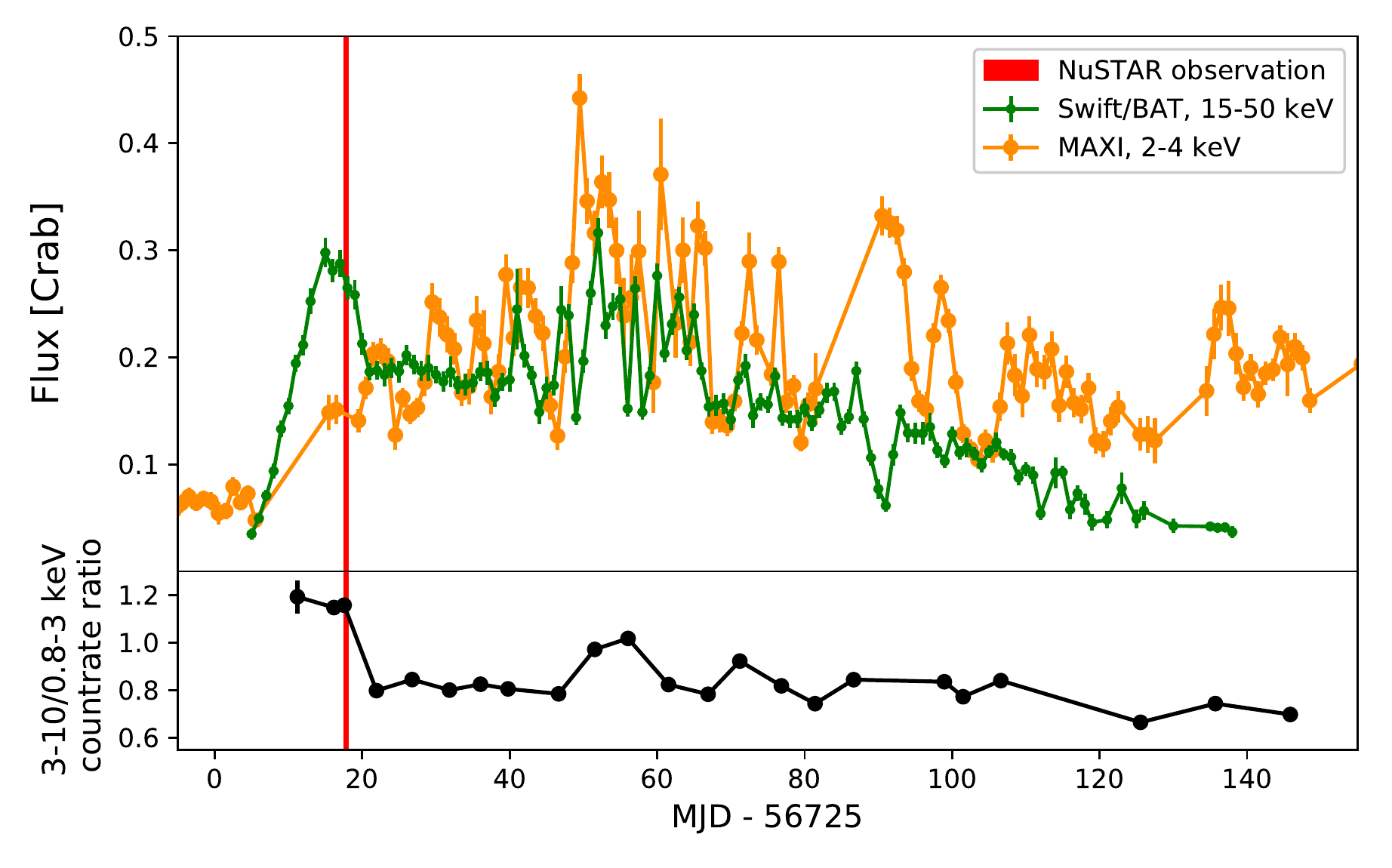}}
\caption{Upper panel: the light curve of \grs\ during the 2014 outburst  as observed by \swiftb\ in the 15--50~keV energy band (green points) and  by the \maxi\ monitor in the 2--4 keV band (orange circles).  
The red vertical line shows the time interval of the \nustar\ observation. 
Bottom panel: evolution of the source spectral hardness during the outburst from the \swiftx\ data.} 
\label{fig:batlc}
\end{figure*}

\citet{borozdin98} found strong spectral evolution throughout the 1996 outburst, consistent with the canonical model -- the outburst starts from LHS, then the soft emission, associated with the optically thick disc starts to dominate, heralding a transition to the HSS. 
Eventually, the source entered the very high state during which a QPO at 5 Hz was detected \citep{borozdin00, 2001MNRAS.328..451W}.

After 18 years of quiescence \grs\ demonstrated another big outburst, a rise of which was detected by \swiftb\ \citep{krimm14_atel} as well as by the \integral\ \citep{filippova14}. 
During this outburst an extensive observing campaign by the \swiftx\ telescope was carried out, together with a single long \nustar\ observation. 
After this outburst the source remained active with a number of repeating mini-outbursts detected \citep{mereminskiy17grs,yan17}.

\section{Observations and data reduction}
\label{sec:data} 

In order to characterize the overall outburst profile we used data from the \swiftb\ transient monitor \citep{krimm13bat} in the hard X-rays (15--50 keV) as well as the data in the soft 2--4 keV band from \maxi\  \citep{matsuoka13maxi}.  
The corresponding reference values for the Crab are 0.22 cts~cm$^{-2}$~s$^{-1}$ (15--50 keV) and 1.67 cts s$^{-1}$ (2--4 keV).

We used the \nustar\ observation (ObsID: 80002018002) performed on 2014 March 26 (MJD 56742) and utilized the \texttt{nuproducts} pipeline to extract photons from the circular region of radius 2\arcmin\ centered at the source in order to produce light curves and the spectra.
We also used public observation by \swiftx\ (target ID: 33203) performed regularly over the rise and peak of the outburst.  
Because the source was bright, all \swiftx\ observations were performed in the windowed mode (WT), allowing to study the timing properties of the source. 
Photons with energies below 0.8 keV and above 10 keV were filtered out. 
The long-term light curves and the spectra were obtained from UK Swift Science Data Centre at the University of Leicester \citep{evans09}
For the spectral fitting we used \texttt{xspec}  v12.9.1p package \citep{arnaud96} and in order to estimate errors we employed Markov chain Monte Carlo method \citep{goodman10_gw}. The element abundances were taken from \cite{wilms00} and the cross-sections from \cite{verner96}. 

\begin{figure*}
\centerline{\includegraphics[scale=0.7]{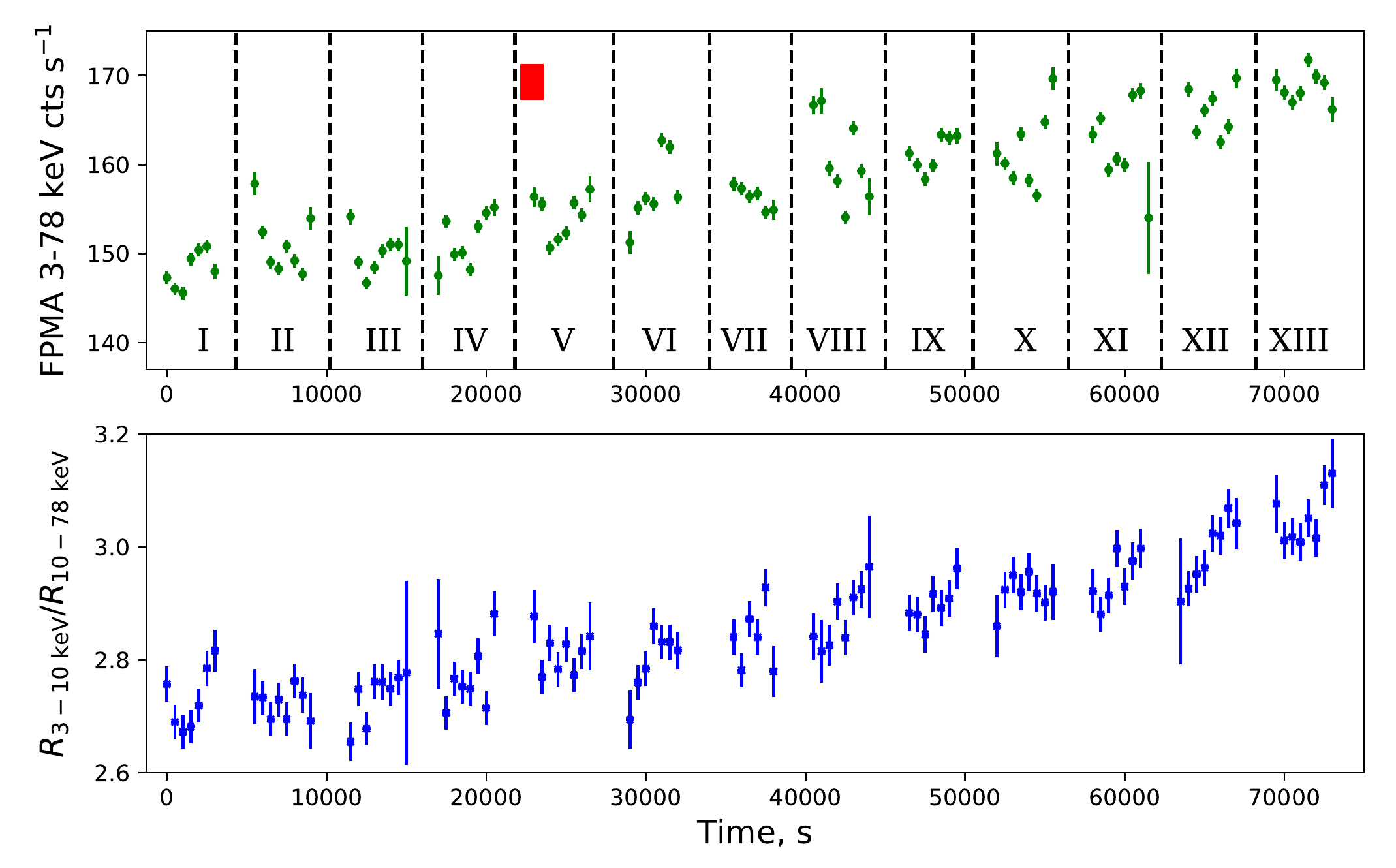}}
\caption{Upper panel: the light curve of \grs\ in the 3--78 keV band as measured by \nustar\ FPMA. 
The thirteen intervals (orbits) of uninterrupted observations are marked with Roman numerals. The red rectangle shows the time of simultaneous \swiftx\ observation (ObsId: 00033203003, second part). Bottom panel: evolution of hardness during observation.
} 
\label{fig:nust_lc}
\end{figure*}

\subsection{Outburst}

The first detection of the source by \swiftb\ occurred at 2014 March 9 (MJD 56725 \citep{krimm14_atel}, which we will use as the zero point). 
An outburst profile in the hard X-rays (15--50~keV) features the fast rise with a tenfold flux increase over ten days (see Fig.~\ref{fig:batlc}), a nearly flat-top peak in the interval 10--15 days from the start followed by an abrupt flux decrease by 30 per cent over the following two days.
After this, the source demonstrated a gradual decline interrupted by a flaring activity at 50--70 d. 
At $\approx 86$ d, a sharp dip occurred in the \swiftb\ light curve. 
After the cease of the outburst, the source has remained active at least until the late fall of 2016 \citep{mereminskiy17grs,yan17} with the flux of about 5--15 mCrab.

Combination of the \maxi\ and \swiftb\ data gives us another insight on the outburst evolution. 
Comparing fluxes in the soft and the hard bands we can see that the soft component lags the hard emission at the beginning of the outburst, but then starts to grow and becomes dominating during the flaring period as well as during the hard dip. 
The bottom panel of Fig.\,\ref{fig:batlc} shows the evolution of the hardness ratio (3--10/0.8--3~keV) measured by the \swiftx. 
The detailed analysis of the spectral evolution during the outburst will be presented elsewhere (Bykov S.~D. et al., in preparation).
The \nustar\ observations \citep{miller15_nust} were carried out right at the transition between the LHS and the HSS, giving us a unique opportunity to study processes that happen during the HIMS.

\section{Spectral properties}
\label{sec:spectra} 

\nustar\ observed \grs\ for nearly 30~ks of the net exposure right after the hard X-ray peak (at $\approx+18$~d, see Fig.~\ref{fig:batlc}). 
Given the 96.9 minute orbital period of \nustar, the observation is divided into 13 orbits separated by the Earth occultations, as shown in Fig.\,\ref{fig:nust_lc}. 
We denoted these orbits with Roman numerals, from I to XIII. 
The source flux increased throughout the observation from $\approx$145 up to $\approx$170 cts s$^{-1}$. 
The spectrum is also altered, with the hardness (defined as a ratio of count rates  $R_{3-10\,\rm keV}/R_{10-78\,\rm keV}$) that had grew monotonically from 2.7 to 3.1 (Fig.\,\ref{fig:nust_lc}).

We used \swiftx\ observation (ObsId: 00033203003, with 1.3 ks exposure) that coincides with the \nustar\ observation, to extend the energy range up to 0.8--78 keV. This allowed us to search for the thermal emission associated with the cold standard disc with $kT \sim 0.1-0.4$~keV, typical for other accreting BHs  \citep[see e.g.][]{miller06b,miller06a,parker15}.
We extracted \swiftx\ spectrum using only zero-grade events, grouped it to have at least 30 counts per bin and added 3 per cent systematic error.\footnote{See XRT CALDB release notes \url{http://www.swift.ac.uk/analysis/xrt/files/SWIFT-XRT-CALDB-09_v16.pdf}} 
Similar grouping were applied to the \nustar\ data and spectra from two \nustar\ modules were fitted simultaneously with free cross-calibration constant between modules. We added 1 per cent systematic error to the \nustar\ data, because it was found that the differences between Crab model spectra and observations is below 2 per cent.\footnote{see \url{https://heasarc.gsfc.nasa.gov/docs/nustar/nustar_obsguide.pdf}} For broad-band spectral fitting we used \nustar\ data in the 4.5--78 keV range, while for fits of solely \nustar\ data we used the full energy range of 3--78 keV.

\begin{figure}
\centerline{\includegraphics[width=\linewidth]{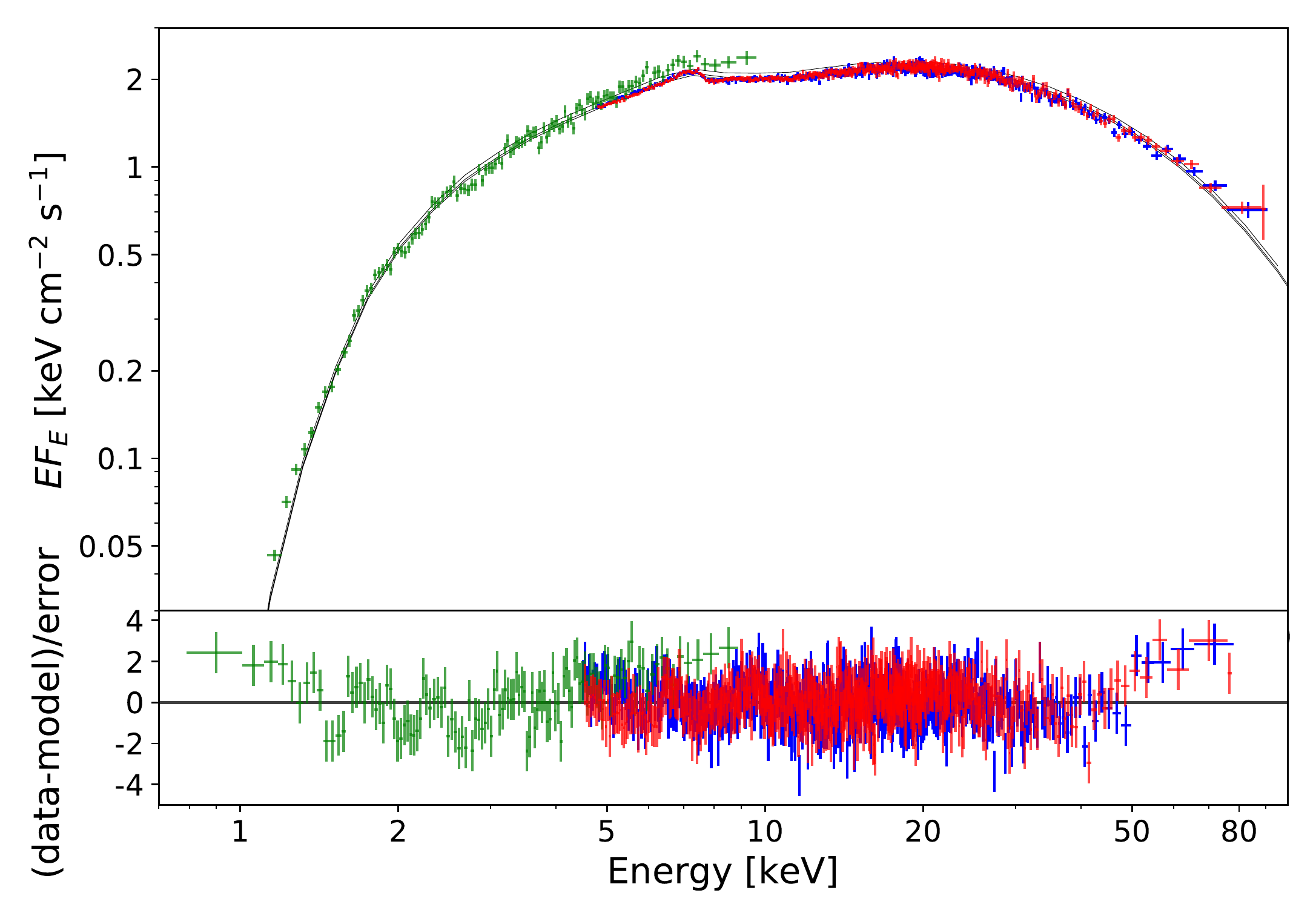}}
\caption{The composite \swiftx\ and \nustar\ spectrum and the best-fitting \textsc{phabs*relxilllp} model. The green, red and blue points correspond to the \swiftx, \nustar\ FPMA and FPMB data, respectively. Data were rebinned for clarity.
} 
\label{fig:spec}
\end{figure}

\subsection{Broad-band average spectrum}
\label{sec:spec}   

Using the same \nustar\ observations \citet{miller15_nust} showed that the average spectrum of \grs\ is well described by the relativistic reflection models such as \textsc{reflionx} \citep{ross05} and \textsc{relxill} \citep{garcia14, dauser14,dauser16} with the accretion disc  reaching  very close to the BH ISCO. 
The disc inner edge radius was estimated as $R_{\rm in} = 5^{+3}_{-4}\, G M/c^{2}$ \citep{miller15_nust}. 
It was also noted that no additional thermal component is needed to describe the energy spectrum, probably due to the low disc temperature and high absorption.
We would like to mention that the obtained spectrum (Fig.~\ref{fig:spec}, see also fig.~2 in \citealt{miller15_nust}) has a very complicated shape, different from the canonical LHS spectrum \citep{ZG04}, when the power law associated with thermal Comptonization extends up to at least 100 keV without a cut-off, as was observed in \grs\ during the failed outburst in 2016 \citep{mereminskiy17grs}.

To fit broad-band average spectrum we applied \textsc{relxilllp} (v1.0.4) spectral model that describes the reflection of emission, produced by a point source located on the rotation axis above the Kerr BH, from the relativistic accretion disc. 
We selected this model for several reasons - first, \citet{miller15_nust} found that it matches well the \nustar\ data. 
Second, during the 1996 outburst the source was detected in the radio, possibly indicating the jet activity and the jet base is often thought to be responsible for this type of ``lamp-post'' geometry. 

Usage of \textsc{phabs*relxilllp} spectral model with fixed absorption column of $N_{\rm H} = 2.15\times10^{22}$~cm$^{-2}$ \citep{fuerst16_gx339} led us to the systematic negative residuals below a few keV. 
Therefore, we left $N_{\rm H}$ free and obtained a value of $2.61\times10^{22}$~cm$^{-2}$. 
No additional soft component is required in order to describe the broad-band spectra.

It should be noted, that \grs\ is known to demonstrate a large dust scattering halo \citep{greiner96}. 
Similar halo was observed during this observation (K.L. Page, priv. comm.). 
It may introduce some bias in the spectrum normalization, which is expected to be up to 15 per cent level. 
Therefore, this discrepancy in the measured absorption column densities could be partially caused by the unaccounted halo emission.

The model describes the data reasonably well with $\chi^{2}_{\rm{red}}\approx1.0$ (Fig.~\ref{fig:spec}). 
The deviations of the \swiftx\ data from the \nustar\ data in the 5--9 keV are probably result of non-simultaneousness of observations.
The best-fitting parameters are listed in Table~\ref{tab:fullfit}. 
The obtained constraints on the accretion disc truncation radius, $R_{\rm{in}} < 7.3\,GM/c^{2}$ (90 per cent confidence limit), and the height of the ``lamp'' above the accretion disc are similar to the values obtained by  \citet{miller15_nust}.
Some discrepancy, seen in the parameters of the accretion disc such as  inclination,  ionization parameter and the iron abundance, can be caused by a broader energy range used in our study and also by usage of the newer version of  \textsc{relxill} model.

The total unabsorbed flux in the 0.1--100 keV band is about $1.3\times10^{-8}$~\ferg\ which translates to the luminosity of $1.0\times10^{38}$~erg~s$^{-1}$ for the distance of 8~kpc. 
Typical luminosity at which BHs transit from a LHS to HIMS is about $0.1 L_{\rm Edd} = 1.2\times10^{37} (M/{\rm M}_{\odot})$ erg s$^{-1}$ \citep{2010MNRAS.403...61D}, therefore one can get a rough estimate for the BH  mass of 8M$_{\odot}$. Although, we should note that there is a significant scatter in this value.

\begin{table}
\noindent
\centering
\caption{Best-fitting parameters of \textsc{phabs*relxilllp} model.}
\label{tab:fullfit}
\centering
\begin{tabular}{|c|c|c|}
\hline\hline
Parameter & Units & Value \\
\hline
$N_{\rm H}$ & $10^{22}$ cm$^{-2}$ &   $2.61^{+0.01}_{-0.02}$ \\   
$h$ & $GM/c^{2}$   &  $26^{+14}_{-5}$ \\
$a$ & $cJ/GM^{2}$    & $0.41^{+0.26}_{-0.59}$   \\
inclination & deg & $8.7^{+1.0}_{-0.1}$ \\
$R_{\rm{in}}$ & $R_{\rm{g}}$ & $5.3^{+2.0}_{-0.5}$ \\ 
$\Gamma$& & $1.36^{+0.02}_{-0.01}$   \\
$\log\xi$ & &  $3.78^{+0.06}_{-0.09}$ \\
$A_{\rm{Fe}}$ &   &  $3.8^{+0.6}_{-0.8}$  \\        
$E_{\rm{cut}}$ & keV    &  $26.1^{+0.6}_{-0.7}$    \\
$R_{\rm{refl}}$ &  &  $0.44^{+0.11}_{-0.04}$    \\
$N_{\rm{FMPA}}$    &  $10^{-2}$    &      $1.32^{+0.12}_{-0.14}$ \\
$C_{\rm{FMPB}}$ &&  $1.016^{+0.003}_{-0.002}$    \\
$C_{Swift/\rm{XRT}}$ &   &   $1.05^{+0.01}_{-0.01}$\\
$\chi^{2}$/dof  &  &   2969/3062 \\      
\hline
\multicolumn{3}{p{5cm}}{Note: errors are 90\% confidence range, estimated from the MCMC set.}
\end{tabular}
\end{table}

\subsection{Spectral evolution and constraints on the disc inner radius}
\label{sec:continuum_evolution}

To get a better view on the evolution of the continuum emission we fitted all individual orbit spectra with the absorbed \textsc{xillver} \citep{garcia13} model, i.e. \textsc{phabs*xillver}. 
This model describes the reflection of the incident radiation from an ionized slab. 
The spectrum of the incident radiation is assumed to be a power law with an exponential cut-off. 
We picked the \textsc{xillver} model over the \textsc{relxilllp} for the analysis of separate orbits because we wanted to describe changes in the continuum emission making no assumptions on the system geometry. 

\begin{figure}
\centerline{\includegraphics[width=\linewidth]{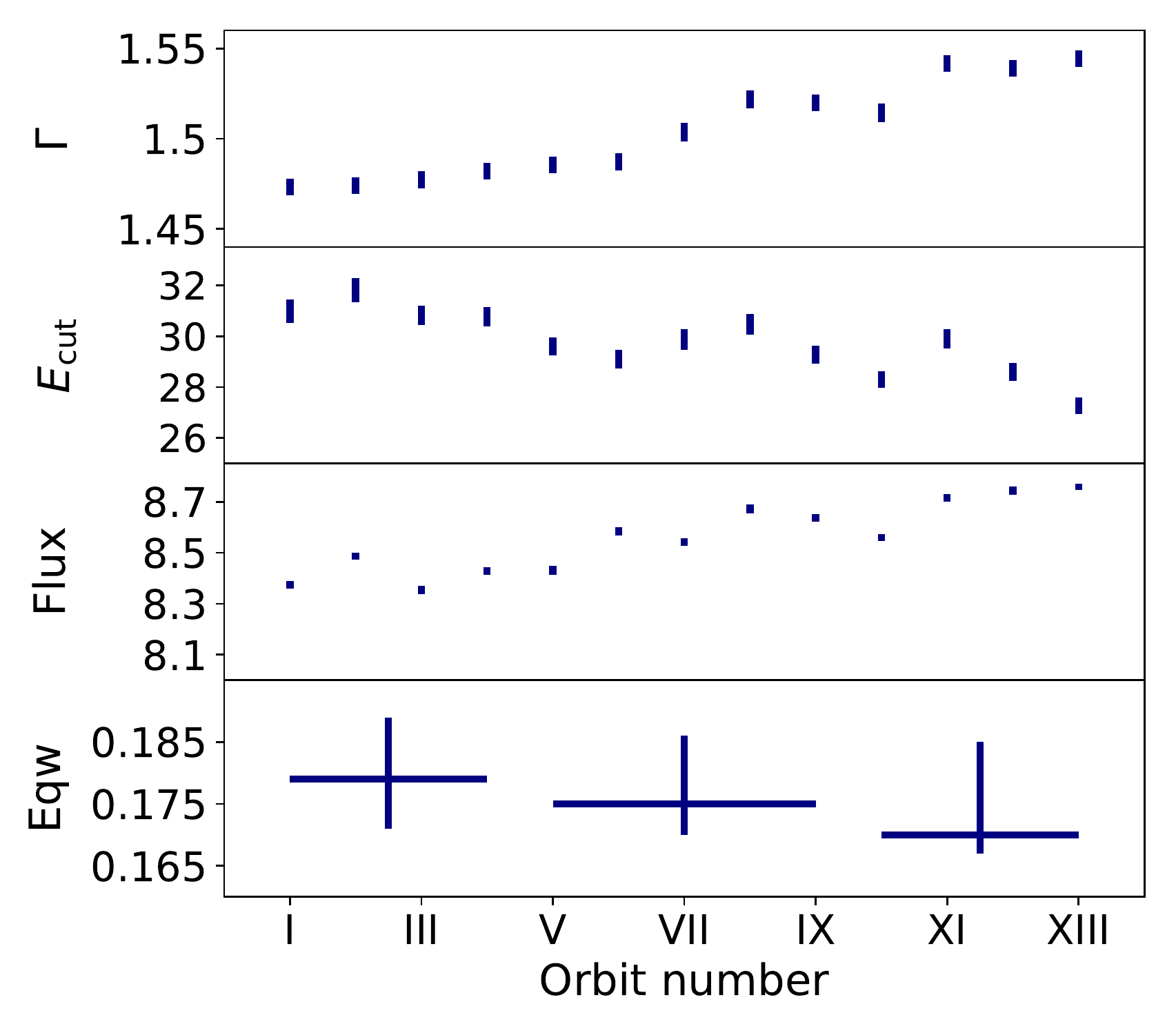}}
\caption{Parameters of the continuum emission of different orbits. From the upper to the lower panel: the \textsc{xillver} power-law slope, the cut-off energy (keV),  the flux in the 3--60~keV band (in units $10^{-9}$\ferg) and the equivalent width of the Fe K$\alpha$ line (keV) obtained with \textsc{phabs$\times$(cutoffpl + gauss)} model (see Section~\ref{sec:continuum_evolution}).
} 
\label{fig:intspe}
\end{figure}

Before fitting, the spectra were grouped in order to have at least 100 counts per bin, channels above 60 keV were ignored, due to the low statistics of high-energy photons. 
Since we used only \nustar\ data for this analysis we fixed the interstellar absorption, the relative iron abundance, the ionization parameter and the inclination at values that were derived from the broad-band spectrum. 
Although \textsc{xillver} has no relativistic broadening of the Fe K$\alpha$ emission line, no significant residuals in the 5--8~keV~range are seen, mainly because of the limited statistics of the individual spectra. 
The resulting fits are of satisfactory quality with $\chi^{2}/$dof$\approx 1.1$. 
The examination of the best-fitting parameters (see Fig.~\ref{fig:intspe}) confirms that the spectrum softens during the observation and the cutoff energy decreases. 

The spectra of individual orbits do not have enough statistics to constrain the changes of the Fe-line profile and, as a consequence, to determine whether the disc inner boundary moves during the observation. 
To increase the statistics, we split the whole observation into three major parts, with the first one consisting of orbits I-IV, the second of  V-IX and the third of X-XIII. 
The \nustar\ 3-78 keV spectra were grouped in order to have at least 100 counts per bin and then fitted with the simple \textsc{phabs*cutoffpl} model, using $N_{\rm H}$ estimated from broad-band spectrum and excluding the data in the 5--10 keV interval.  

\begin{figure}
\centerline{\includegraphics[width=\linewidth]{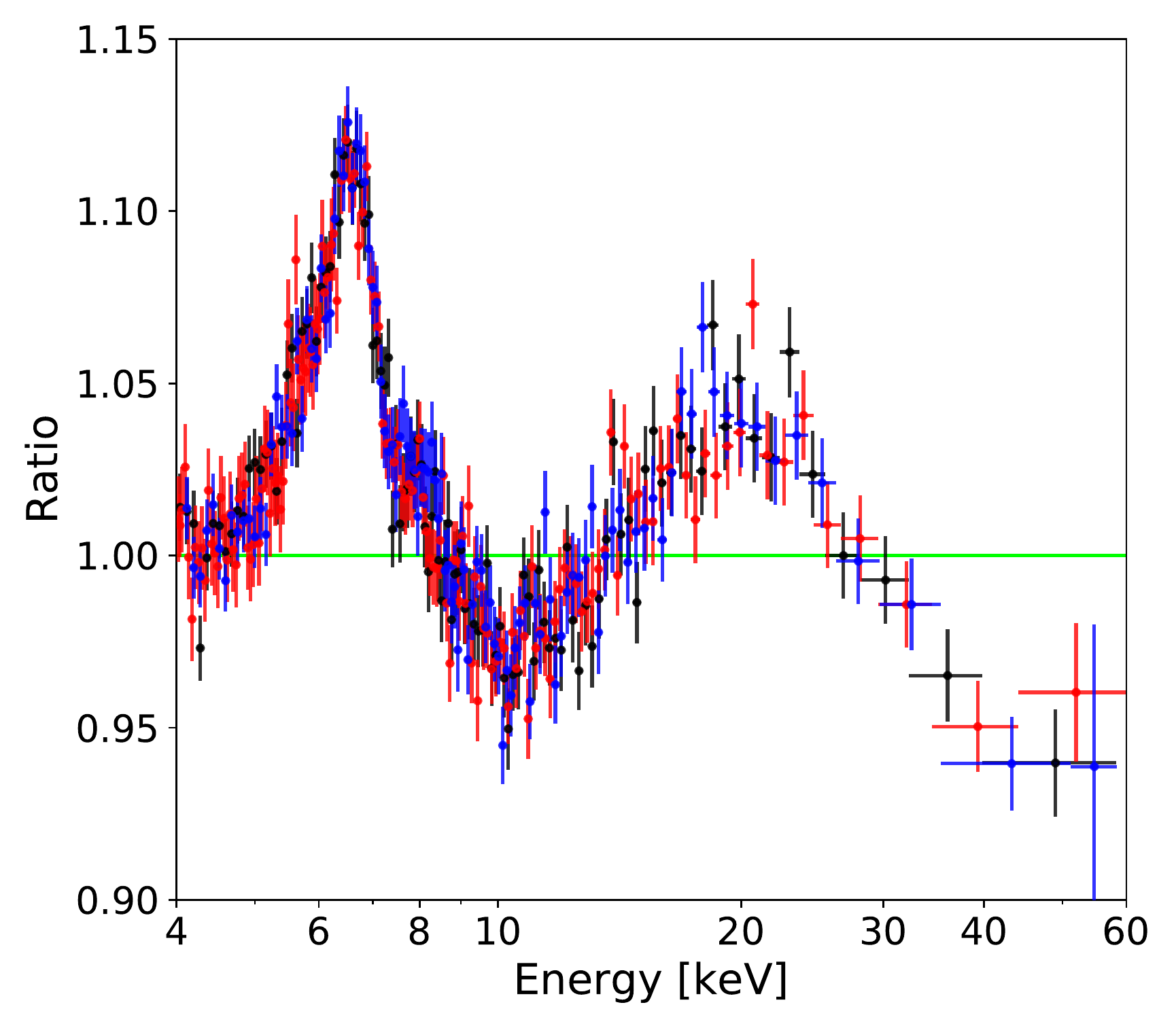}}
\caption{Ratio of the \nustar\ FMPA spectra to the best-fitting \textsc{phabs*cutoffpl} model. 
The data from the orbits I-IV,  V-IX and X-XIII are shown with black, red  and blue crosses, respectively.} 
\label{fig:ratios}
\end{figure}

The ratio of the data to this best-fitting model  (Fig.~\ref{fig:ratios}) demonstrates two prominent features, a broadened iron line at 5--8 keV and Compton hump around 20 keV. 
The equivalent width of the line was estimated using a continuum model of an absorbed cut-off power law and a Gaussian line  and ignoring the data in 10--30~keV range (to neglect  the contribution of the Compton-hump). 
The equivalent width of the line was about 0.175 keV and remained constant during the observation within the errors.
Therefore we can conclude, that there were no drastic change in position of the inner disc boundary during the observation.

\section{Timing analysis} 
\label{sec:timing}

Properties of X-ray binaries in the time domain can be described with different metrics. 
Variability properties of different types of X-ray binary systems are usually described in terms of the power spectrum.
The power spectrum of the BH systems in LHS/HIMS state can be described typically as a combination of a band-limited noise and one or few narrow Lorentzian functions, representing QPOs \citep[see, e.g.,][]{1972ApJ...174L..35T, 1990A&A...227L..33B, homan05}. 
Properties of these components and correlations between them, in principle, may be used to discriminate between different models of the formation of the X-ray emission in BH systems. 
 
Although the power spectra analysis is by far the most popular method to study physical properties of the accretion flow, more sophisticated methods are applied as well. 
As an example of such methods we can name the frequency-resolved spectroscopy, autocorrelation function and cross-correlation function and the time-lag estimation between different energy bands.
The frequency-resolved spectroscopy \citep{2000MNRAS.316..923G} was used to determine the spectra of the component responsible for the source variability.
The autocorrelation function and the cross-correlation function \citep[see, e.g.,][]{2000ApJ...537L.107M, 2014SSRv..183...61P} were used to constrain geometrical size of the regions responsible for the emission in a particular energy band and also associate the emission in different energy bands with specific spectral components.
The time-lags between the soft and hard emission \citep{1997ApJ...474L..43V,1999ApJ...517..355N} also allow to constrain the geometrical size of the accretion flow and sift some of the noise formation models. 

\subsection{Power spectrum}

\begin{figure*}
\includegraphics[width=0.495\textwidth]{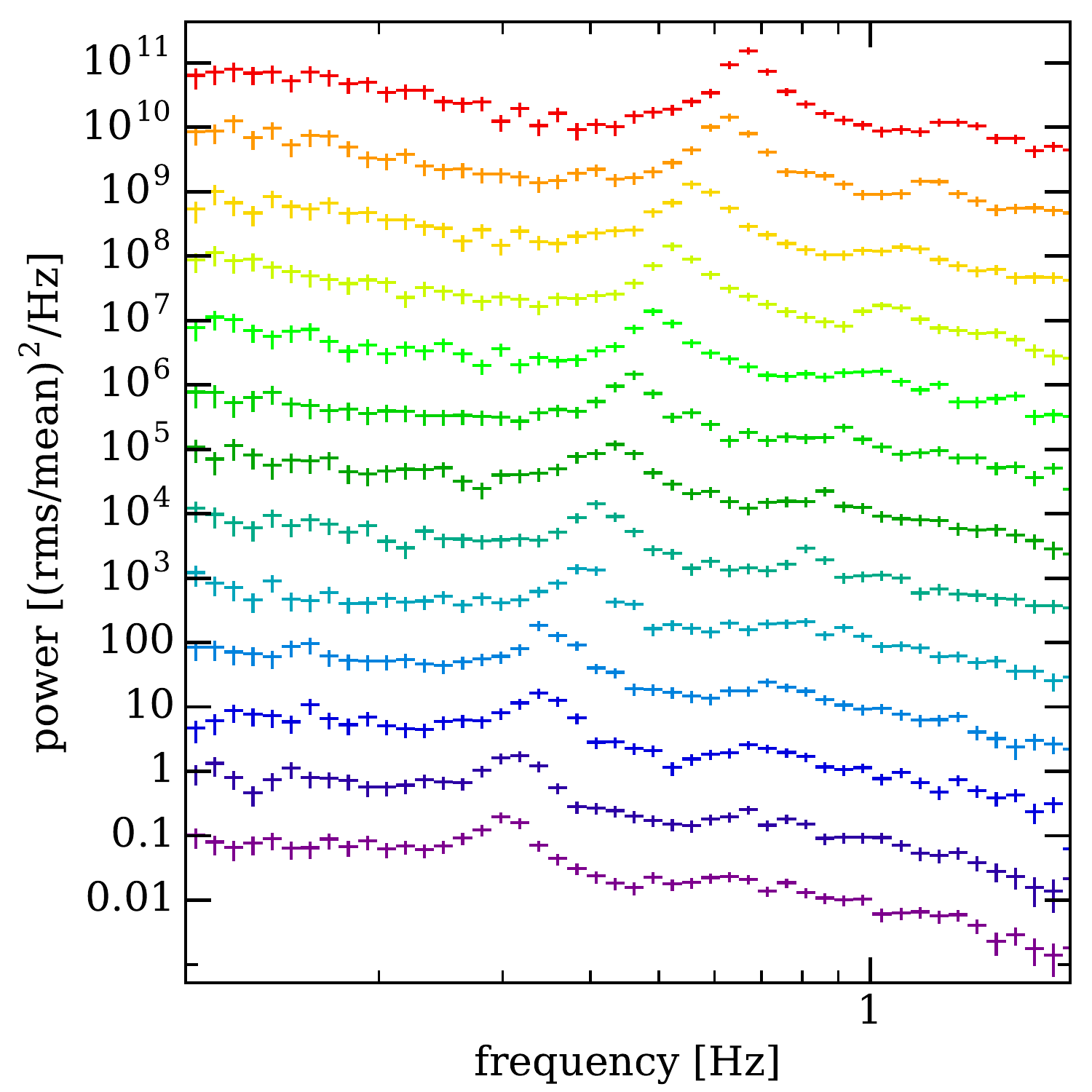}
\includegraphics[width=0.495\textwidth]{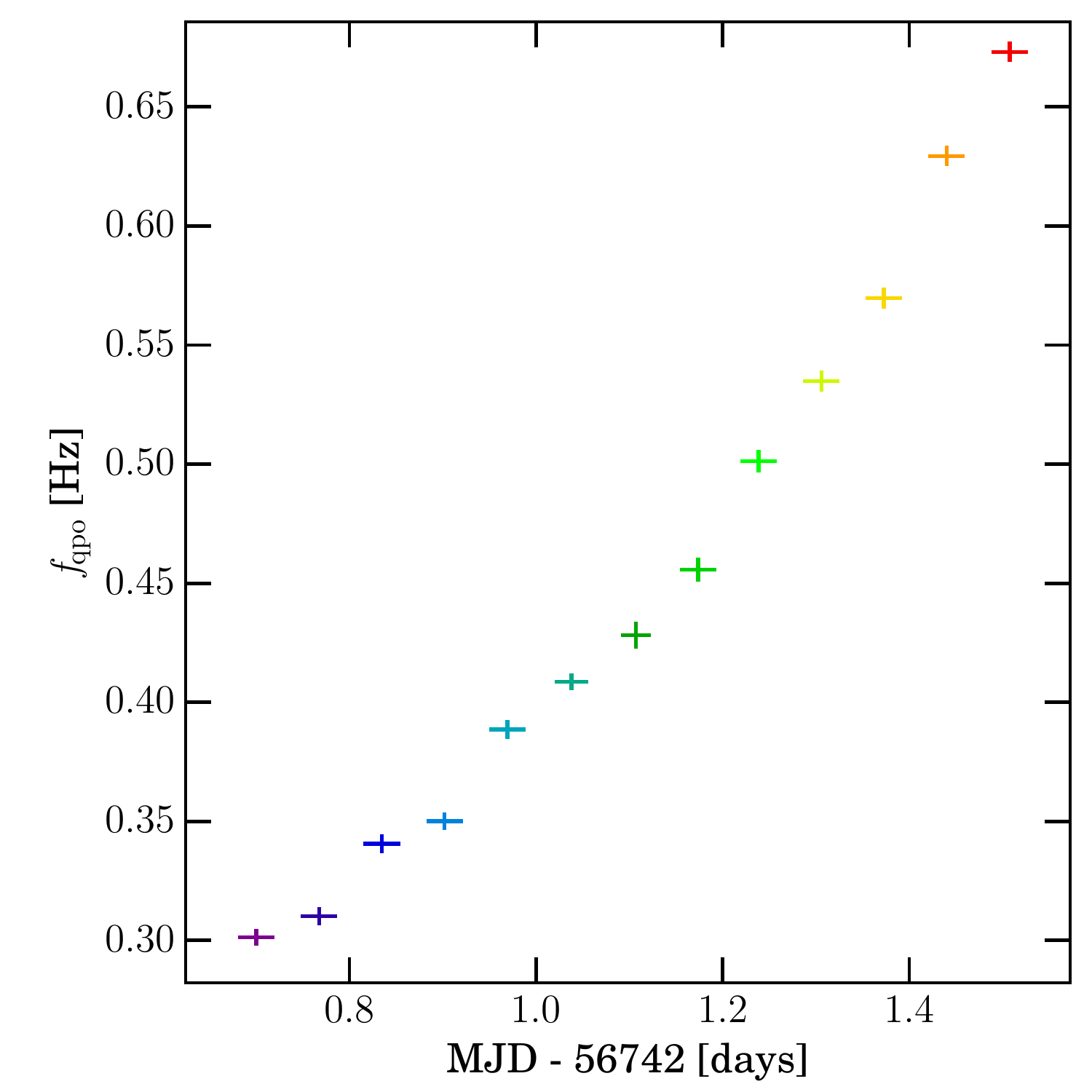}
\caption{Left: The power spectra of \grs\ obtained in different \nustar\ orbits (I to XIII, from bottom to top). 
Each following power spectrum is multiplied by 10 for clarity.
Right: The time evolution of the QPO first harmonic centroid frequency. 
}
        \label{fig:qpo}
\end{figure*}

As it was mentioned above, we split \nustar\ observation of \grs\ into 13 continuous intervals (orbits) separated by $\sim0.7$~hr gaps when the source was occulted by the Earth. 
The continuous orbits have a duration of about 3~ks (see Table~\ref{tab:timing}).
Because the \nustar\ detectors operate in the photon counting mode, the light curve can be constructed with the time resolution down to 2~$\mu$s.
For our analysis we extracted light curves with the 0.01~s resolution in several energy bands (3--78, 3--5, 5--8, 8--15, 15--78~keV), which allows us to examine the frequency range of 0.003--50~Hz.
This frequency band usually contains low-frequency QPOs and the broad-band noise \citep{wijnands99}.
We calculated the power spectra for each orbit using the light curves in the 3--78~keV energy band.
All power spectra are similar (see Fig.~\ref{fig:qpo}): a plateau ($P(f)\propto$const) at low frequencies, transforming  to the power law with a slope of $\rho\approx-1.6$..$-2.0$ at the frequency of $\approx0.1$~Hz and the Poisson noise plateau at the frequencies above a few Hz. 
A prominent QPO at the frequencies of 0.3--0.7~Hz and its second harmonic are present as well.

In order to quantitatively characterize properties of the broad-band noise and QPOs we approximate each obtained power spectra with the following analytical function:
\begin{eqnarray}\label{eq:ps_model}
P(f)  & = & N \left[1 + (f/f_{\rm lb})^4\right]^{\alpha} 
+ \frac{s_1}{(f - f_{\rm QPO})^2 + (f_{\rm QPO}/\quality)^2} \nonumber \\
& +& \frac{s_2}{(f - 2f_{QPO})^2 + (2f_{\rm QPO}/\quality)^2} + P_{\rm Poiss}, 
        \label{eq:complex_fit}
\end{eqnarray}
where $f_{\rm lb}$ is the broad noise break frequency, $f_{\rm QPO}$ and $\quality$ are the centroid of the QPO and its quality, respectively, $N$ is the scale factor, defining broad-band noise total power, $P_{\rm Poiss}$ represents the mean power of the variations caused by the counting statistics and dumped by the dead-time.
Here the first component represents the plateau with the break, the following two components describe QPO's first and second harmonics and the last component represents the Poisson noise.
We assume that the quality of the first and second harmonic of the QPO is equal.
In the following we refer to this model as standard.

In order to determine properly all parameters one has to know the shape and normalization of the Poisson noise component, which depend on the count rate and the dead-time and in principle can be described with the analytical functions \citep[see, e.g.,][]{1994A&A...287...73V, 1995ApJ...449..930Z}.
\nustar\ detectors are subject to a non-paralyzable dead-time with the characteristic time-scale of $\tau \approx 2.5$~ms \citep{2015ApJ...800..109B}.
For \grs\ the effects from the dead-time can be observed in the power spectra at frequencies above 20~Hz.
\citet{2015ApJ...800..109B} noted that the \nustar\ dead-time has a complex dependence on the energy of registered photons, and therefore it is hard to create an analytical model for the power spectrum of the Poisson noise. 

To avoid this problem \citet{2015ApJ...800..109B} proposed to use a cross-spectrum (or shortly cospectrum) for the analysis of the \nustar\ data instead of the power spectrum. 
The authors define the cospectrum as a real part of the cross product of the Fourier transforms of light curves obtained from the two \nustar\ detectors 
\begin{equation}
        \label{eq: ps_estimation}
        P(f) \approx {\Re\langle F_{\rm FPMA}^{*}(f)F_{\rm FPMB}(f)\rangle} , 
\end{equation}
where $P(f)$ is an estimation of the source intrinsic variability power spectrum, $F_{\rm FPMA[B]}$ is the Fourier transform of a light curve from FPMA[B] module and the asterisk stands for the complex conjugation. 
This approach is based on the following assumptions: signals, produced by the observed source on two detectors, are identical and have no time lag, therefore their Fourier transforms are also identical and have a zero phase shift; in contrast, signals independent for two detectors (like counting statistics) have random phase shifts.  
As a consequence, for independent signals the average real part of the cross product tends to zero, i.e. the Poisson noise is eliminated.

\citet{2018ApJS..236...13H} showed that the cospectrum  in each frequency bin is distributed with the Laplace probability density function (PDF), if it is derived from two normally distributed random independent series (see eq.~14 in the above paper):
\begin{equation}
        p(Co|0, \sigma_x \sigma_y) = \frac{1}{\sigma_x \sigma_y} \exp{\left(\frac{-\lvert{Co}\rvert}{\sigma_x \sigma_y} \right)} ,
\end{equation}
where $Co$ is the cospectrum of two incoherent series measured in the specified frequency channel and $\sigma_x$, $\sigma_y$  are standard deviations of the two normal distributions.
These values ($\sigma_x$, $\sigma_y$) are equal to the square root of the power spectra in the corresponding frequency channel for each time series.
If signals, used for the cospectrum estimation, have identical power spectra then $\sigma_x = \sigma_y \approx |F_{\rm FPMA}|$.
We, therefore, see that to determine proper likelihood function which can be used to approximate cospectra with analytical functions, one still has to know the Poisson noise level.
It is also worth noting, that both the source and the background count rates are usually slightly different for the two \nustar\ modules making amplitudes and shapes of the counting-statistic in the power spectrum not equal. 
Taking all these arguments into account we decided to use the standard power spectrum analysis to estimate properties of the source intrinsic variability.

Because we used a relatively large time binning (10 ms) to extract the light curves from the \nustar\ data and considered variability at frequencies below 10~Hz (where signal-to-noise ratio is sufficient), we assumed that the  dead-time only lowers the constant Poisson noise level by a factor $(1 - 2\nu \tau_{\rm d})$, where $\nu$ is the total count rate for a detector and $\tau_{\rm d}$ is the dead time \citep{1994A&A...287...73V, 1995ApJ...449..930Z}. 
As the dead-time is not constant with the energy and we calculate the power spectra for light curves obtained in different energy bands, we determined the modified Poisson level for each extracted data set separately.

\begin{figure}
        \includegraphics[width=\columnwidth]{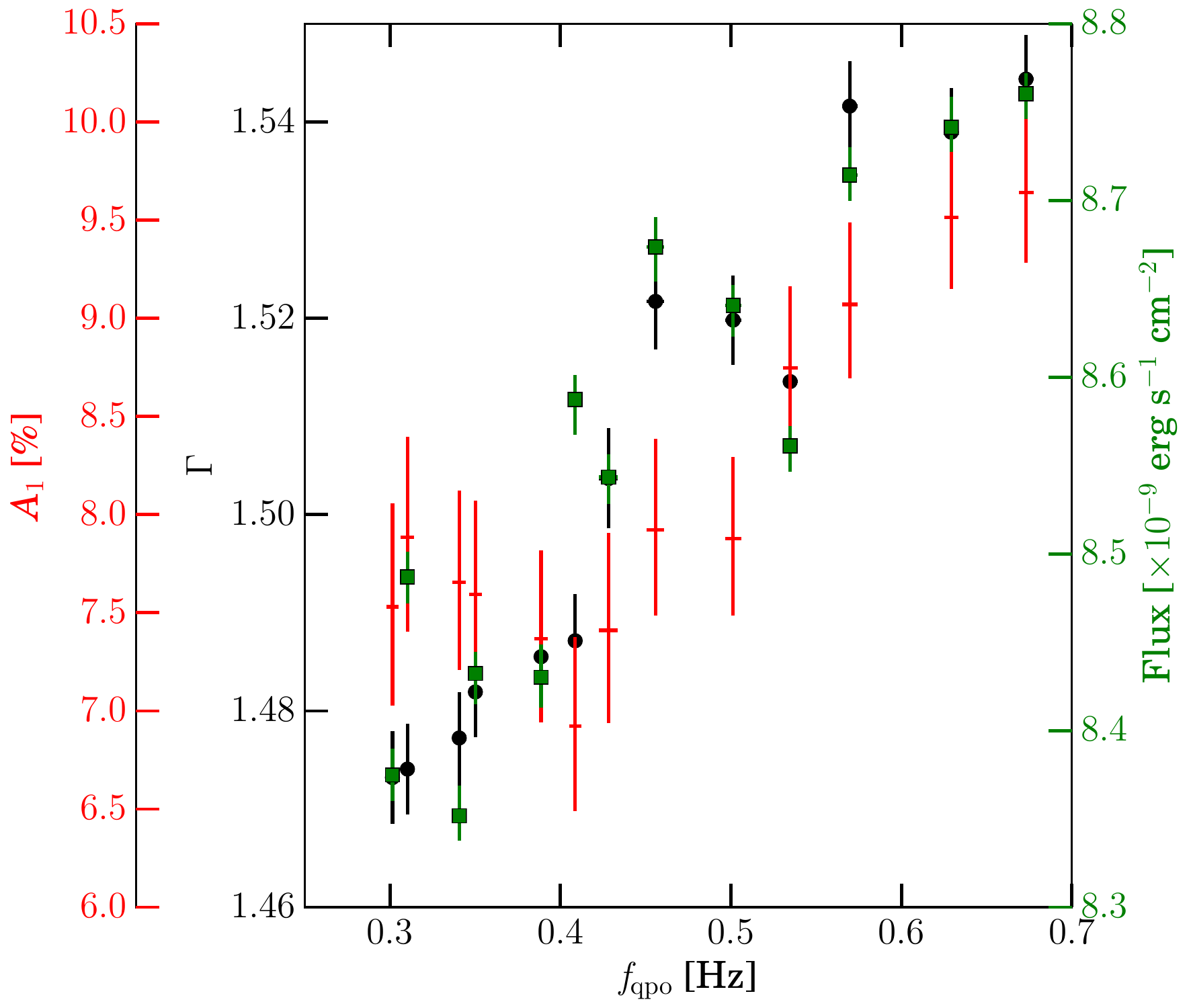}
        \caption{Dependence of the  spectral photon index (black circles with crosses), the QPO amplitude (red crosses) and the source flux in the 3--60~keV band (green rectangles with crosses) on the QPO centroid frequency.
        }
        \label{fig:qpo_gamma}
\end{figure}

We did not consider any high frequency QPOs because for their centroid frequencies of 100--400~Hz, amplitudes of $\approx10$ per cent and quality $\quality\approx$2--10 they would be indiscernible from the Poisson noise with the given count rates and duration of the observation.
We found that the QPO frequency evolves with time (Table~\ref{tab:timing} and Fig.~\ref{fig:qpo}).
It correlates with the \nustar\ flux and the photon index (Fig.~\ref{fig:qpo_gamma}), similar to many other BH and neutron star binary systems \citep[see, e.g.,][]{2000ApJ...531..537S,Revnivtsev01,vignarca03, 2004ApJ...612..988T,Ibragimov05,Gilfanov10,
2014ApJ...789..100S, fuerst16,mereminskiy18_maxi}. 
The QPO amplitude remained stable during the first half of the observation, and started to grow in the second part.

\begin{table*}
\noindent
\centering
\caption{Evolution of the Fourier and energy spectrum properties through the \nustar\ observation in the 3--78~keV energy band.}
\label{tab:timing}
\centering
\begingroup
\setlength{\tabcolsep}{4.pt}
\begin{tabular}{|c|c|c|c|c|c|c|c|c|c|c|c|}
\hline\hline
Orbit    & $T_{\rm start}$  & Exposure & $f_{\rm br}$ & $f_{\rm QPO}$ & $\quality$  & $A_{1}$ & $A_{2}$ & rms & $\Gamma$ & $E_{\rm cut}$ & $\chi^2$ (d.o.f.) \\
      ~      &  MJD   &  s  & $10^{-2}$ Hz &  Hz &   & \%  &  \%  & \% & ~ & keV & ~  \\
\hline
I           & 56742.68      & 3386      & $8.9_{-2.3}^{+2.2}$  & $0.30\pm0.01$ & $15_{-3}^{+5}$ & $7.5_{-1.0}^{+0.9}$ & $2.7_{-0.9}^{+0.8}$ & $26\pm1$ & $1.473\pm0.005$ & $31.0_{-0.5}^{+0.4}$ & 4238 (4192) \\
II & 56742.75 & 3388 & $7.8_{-1.8}^{+2.0}$ & $0.31\pm0.01$ & $13\pm3$ & $7.9_{-0.9}^{+1.0}$ & $2.7_{-0.9}^{+0.8}$ & $26\pm1$ & $1.474\pm0.005$ & $31.9_{-0.6}^{+0.4}$ & 4132 (4194)\\
III & 56742.82 & 3392 & $8.2_{-1.9}^{+2.4}$ & $0.34\pm0.01$ & $12_{-2}^{+3}$ & $7.7_{-0.8}^{+0.9}$ & $3.9_{-0.8}^{+0.9}$ & $26\pm1$ & $1.477\pm0.005$ & $30.8\pm0.4$ & 4364 (4226) \\
IV & 56742.88 & 3389 & $8.3_{-1.8}^{+2.0}$ & $0.35\pm0.01$ & $15_{-3}^{+4}$ & $7.6\pm0.9$ & $3.2_{-0.8}^{+0.7}$ & $26\pm1$ & $1.481\pm0.005$ & $30.7_{-0.3}^{+0.4}$ & 4254 (4215) \\
V & 56742.95 & 3389 & $6.9_{-1.4}^{+1.6}$ & $0.39\pm0.01$ & $13_{-3}^{+4}$ & $7.4\pm0.8$ & $4.3_{-0.9}^{+0.8}$ & $26\pm1$ & $1.486\pm0.005$ & $29.6\pm0.4$ & 4272 (4206) \\
VI & 56743.02 & 3136 & $7.5_{-1.5}^{+1.9}$ & $0.41\pm0.01$ & $17_{-3}^{+5}$ & $6.9_{-0.8}^{+0.9}$ & $3.8_{-0.8}^{+0.7}$ & $26\pm1$ & $1.487\pm0.005$ & $29.1\pm0.4$ & 4128 (3885) \\
VII & 56743.09 & 2771 & $9.7_{-2.2}^{+2.7}$ & $0.43\pm0.01$ & $12_{-2}^{+3}$ & $7.4\pm0.9$ & $3.6\pm0.9$ & $26_{-1}^{+2}$ & $1.503\pm0.005$ & $29.9\pm0.4$ & 3334 (3393) \\
VIII & 56743.15 & 3387 & $5.8_{-1.5}^{+1.4}$ & $0.46\pm0.01$ & $11_{-2}^{+3}$ & $7.9\pm0.9$ & $4.2_{-0.8}^{+0.9}$ & $27\pm2$ & $1.521\pm0.005$ & $30.5\pm0.4$ & 3773 (3632) \\
IX & 56743.22 & 3392 & $7.1_{-1.4}^{+1.6}$ & $0.50\pm0.01$ & $12_{-2}^{+4}$ & $7.9\pm0.8$ & $4.3\pm0.8$ & $26\pm1$ & $1.519\pm0.005$ & $29.3_{-0.4}^{+0.3}$ & 4070 (4212) \\
X & 56743.29 & 3390 & $7.0_{-1.6}^{+1.7}$ & $0.53\pm0.01$ & $13_{-2}^{+3}$ & $8.7\pm0.7$ & $4.5_{-0.8}^{+0.7}$ & $25\pm1$ & $1.513\pm0.005$ & $28.3\pm0.3$ & 4199 (4197)\\
XI & 56743.35 & 3382 & $6.7\pm1.5$ & $0.57\pm0.01$ & $13\pm3$ & $9.1_{-0.7}^{+0.8}$ & $4.0\pm0.8$ & $25\pm1$ & $1.542_{-0.005}^{+0.004}$ & $29.9\pm0.4$ & 4114 (4083) \\
XII & 56743.42 & 3386 & $6.7_{-1.4}^{+1.8}$ & $0.63\pm0.01$ & $14_{-2}^{+3}$ & $9.5_{-0.7}^{+0.8}$ & $4.4\pm0.7$ & $26_{-1}^{+2}$ & $1.539\pm0.004$ & $28.6\pm0.3$ & 4281 (4127) \\
XIII & 56743.49 & 3391 & $7.5_{-1.5}^{+1.7}$ & $0.67\pm0.01$ & $15\pm3$ & $9.6\pm0.7$ & $4.2\pm0.8$ & $25_{-1}^{+2}$ & $1.544\pm0.004$ & $27.3\pm0.3$ & 4197 (4226)\\
\hline
\multicolumn{12}{p{\textwidth}}{Note: $f_{\rm br}$ is the broad-band noise break frequency, $f_{\rm QPO}$ is the QPO centroid frequency, $\quality$ is the QPO first harmonic quality (ratio between the QPO peak width to its centroid frequency), $A_{1}$ and $A_{2}$  are the total powers in the QPO first and second harmonics, respectively (we accepted that the QPO amplitude related with the model (Eq.~\ref{eq:ps_model}) parameters with the relation $A = 100\sqrt{{\pi s Q}/{2 f_{\rm QPO}}}$), rms is the total amplitude of the source variability, $\Gamma$ is the power-law photon index, $E_{\rm cut}$ is the cut-off energy. Parameters $\Gamma$ and $E_{\rm cut}$ were obtained from the spectra of individual orbits using \textsc{xillver} model (see Section~\ref{sec:continuum_evolution}),  the $\chi^2$ column describes the quality of the fits to the power spectra. Because the original power spectrum data points follow the $\chi^2_2$ rather than normal distribution, 
we rebinned the original power spectra by combining $n=16$ neighbouring data points to use $\chi^2$ criterion. We estimate the dispersion of each obtained data point as the model value in the centre of a new bin divided by $\sqrt{n}$}.
\end{tabular}
\endgroup
\end{table*}

\begin{figure}
\includegraphics[width=\columnwidth]{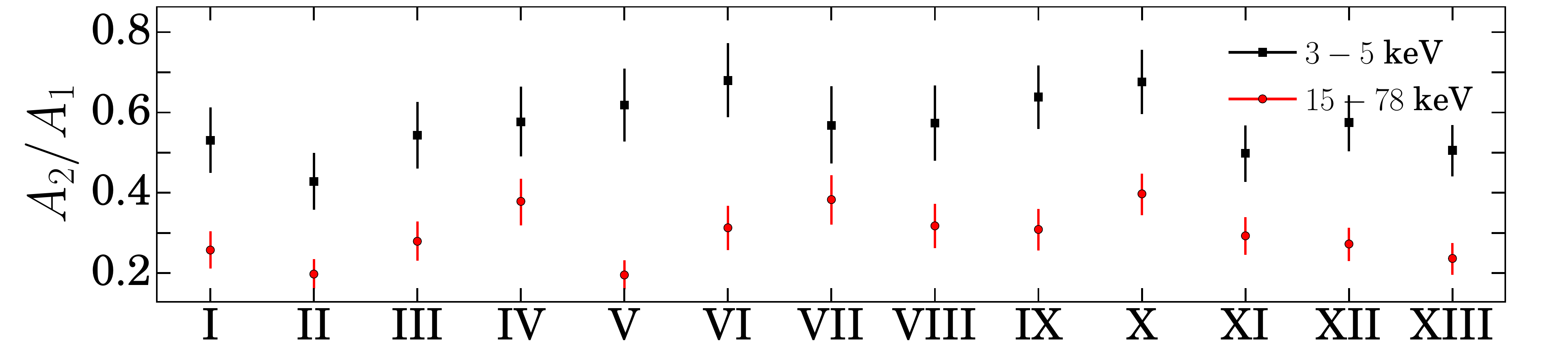}
\includegraphics[width=\columnwidth]{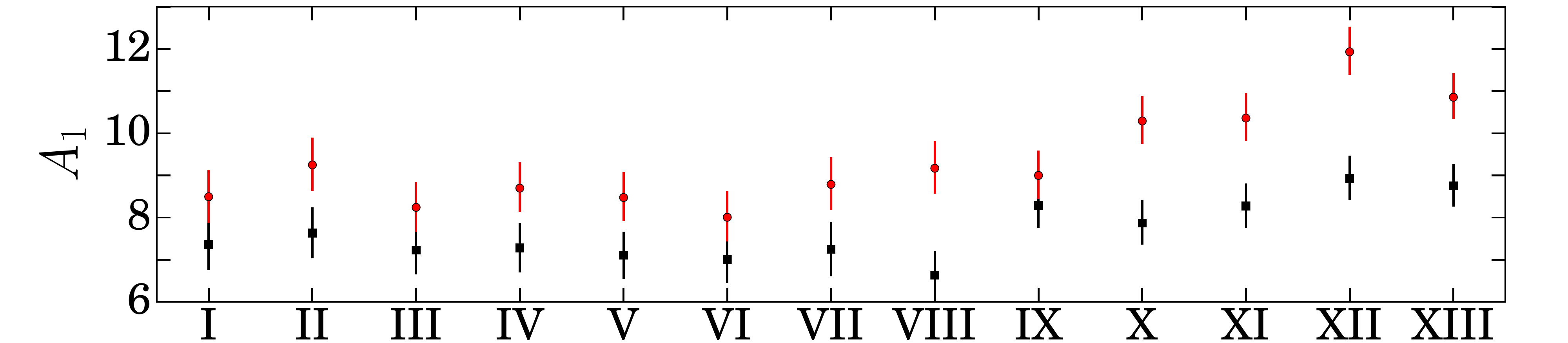}
\caption{Upper panel: the amplitudes of the QPO harmonic in two energy bands: 3--5~keV(black squares) and 15--78~keV(red circle).
Lower panel: the ratio between the powers in the second and first harmonics in the same energy bands.}
        \label{fig:qpo_ratio}
\end{figure}

We also inspected the power spectra in the soft (3--5~keV) and hard (15--78~keV) energy bands and found that the QPO amplitude is smaller in the soft band, while the amplitude of its harmonic is larger.
The ratios of the power in the QPO second and first harmonics for hard and soft energy bands are presented in Fig.~\ref{fig:qpo_ratio}.

If the QPO's first and second harmonics are coherent (Fourier signals of the QPO harmonics have conserving phase shifts between each other) on some time-scales, they correspond to a particular pulse profile in time domain \citep{2015MNRAS.446.3516I}. 
From the changing ratio between the QPO and its harmonics it follows that the QPO pulse profile changes with energy. 
Following \citet{2015MNRAS.446.3516I} we tried to extract QPO profile, however, no significant coherence between two harmonics was detected above the noise level.
It indicates that the pulse profile was not stable during the observation, in contrast with the result obtained by \citet{2015MNRAS.446.3516I} for GRS~1915+105 with {\it RXTE}.

\begin{figure}
\includegraphics[width=\columnwidth]{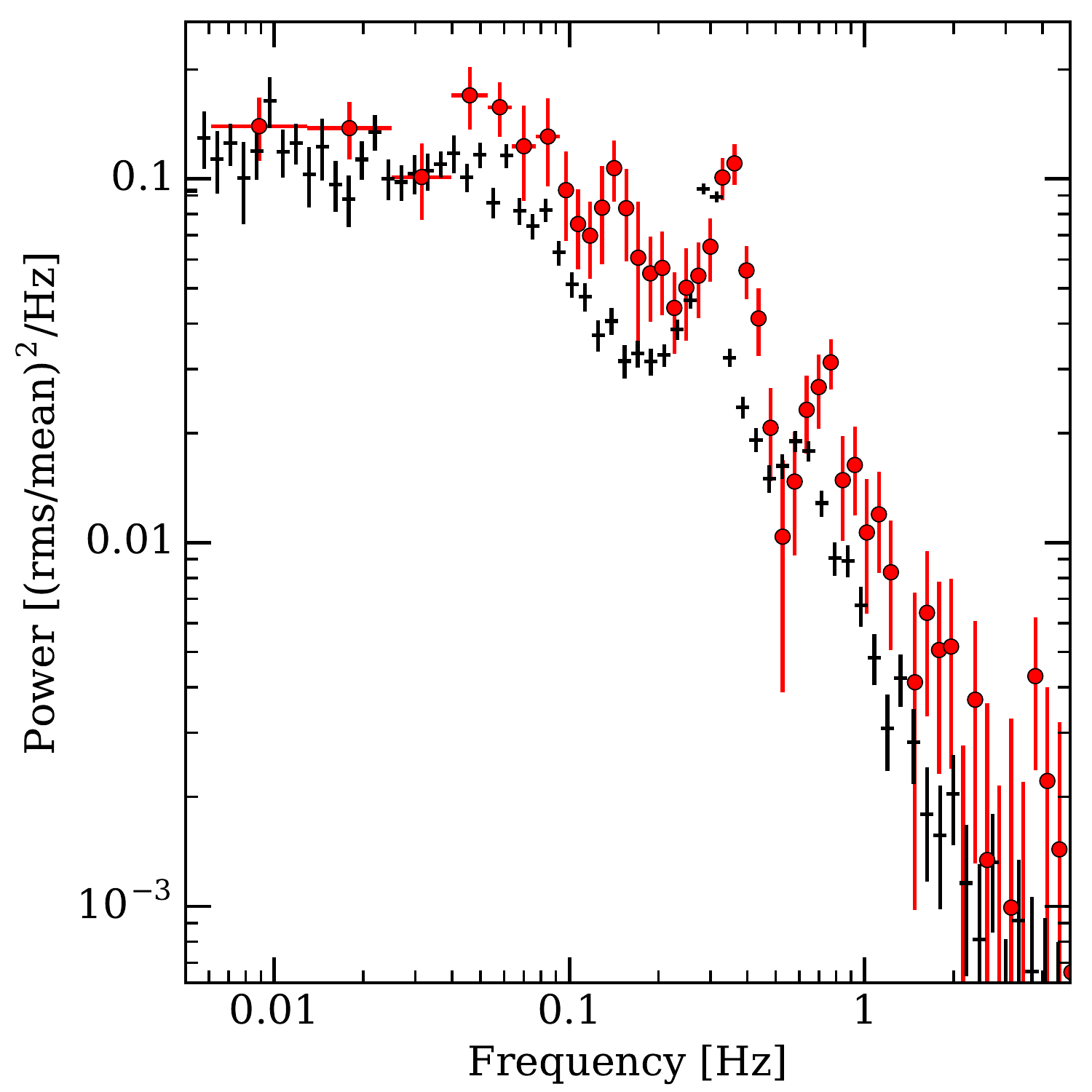}
\caption{Cross-spectrum of the observations, obtained by scaling frequencies to conserve the QPO position.
Black crosses correspond to the data from all orbits, while the red circles  with  crosses are for the orbit IV in which the QPO subharmonic was most prominent.}
\label{fig:cospec_tracked}
\end{figure}

In some orbits a subharmonic, centered approximately at 1/2 of the QPO centroid frequency is clearly observed in the cospectra (see examples in Fig.~\ref{fig:cospec_tracked}, red circles).
In order to detect QPO with the better significance, we stacked several cospectra, the frequencies of each cospectrum were scaled to conserve the QPO centroid at 0.3~Hz.
This way obtained ``tracked'' cospectrum is presented in Fig.~\ref{fig:cospec_tracked}.
The subharmonics seems to roam around half the QPO frequency, therefore it is not seen in the tracked cospectrum.

We should  note that changes in the QPO centroid position during each orbit may contribute to the measured quality factor because of a relatively large rate of QPO frequency variations,  $d{f}_{\rm QPO}/dt \approx (5.0$--$6.5)\times10^{-6}$~Hz~s$^{-1}$, and a relatively short exposure during separate orbits, $\approx3000$~s.
Movement of the QPO frequency during individual orbit results in the measured quality $\quality = d{f}_{\rm QPO}/dt \times ( t_{\rm obs}/f_{\rm qpo} ) \approx 17$, even if its real value is much higher.

To check the influence of this effect on the quality measurements we introduced another model to fit the power spectra, which takes into account the movement of the QPO frequency. 
In this model we fitted simultaneously multiple power spectra obtained from the shorter segments inside each orbit. 
We assumed that the QPO amplitude and quality are the same for all power spectra, but the QPO centroid frequency linearly evolves with time.
The obtained quality factors, with the median value $\quality=14.3$,  are in good agreement with those obtained with a simpler model.

\subsection{Coherence}

\defcitealias{1997ApJ...474L..43V}{VN97}

\citet{1997ApJ...474L..43V} (hereafter \citetalias{1997ApJ...474L..43V}) suggested to use coherence between different energy bands in order to obtain additional information about the source variability. 
The coherence measures the similarity between two signals and can be calculated with the following expression:
\begin{equation}
        C(f) = \frac{\lvert{ \langle F_{\rm s}^*(f)F_{\rm h}(f) }\rangle \rvert^2 - P_{\rm unc}^2}{P_{\rm s}(f) P_{\rm h}(f)} , 
    \label{eq:nowak_coh}
\end{equation}
where $F_{\rm h}(f)$ and $F_{\rm s}(f)$ are Fourier transforms of the observed time series in the hard and soft bands, respectively, $P_{\rm h}(f)$ and $P_{\rm s}(f)$ are the estimations of their power density spectra,  
$P_{\rm unc}^2$ is the product of the power in the uncorrelated noise components divided by the number of used series (which mostly determined by Poisson statistics noise, see \citetalias{1997ApJ...474L..43V}). 
Because the coherence is estimated using a mean product of the Fourier transforms it should be computed for a number of independent time series, therefore we separated each of the available uninterrupted orbits into  several shorter pieces, 82~s long each.  

Different models for generation of the XRBs variability suggest that the signals in two energy bands can be partially independent, while the shape of the power spectra is conserved.
In many sources the coherence between soft and hard X-ray bands is close to unity \citep{1999ApJ...517..355N, wijnands99}.
However there are also indications on a more complex picture of the coherence in some states of different systems, e.g. dip in the coherence at 0.03~Hz frequency, observed in GRS1915+105 \citep[][]{2003ApJ...584L..23J}, decreasing of the coherence between particular energy bands in GX 339--4 \citep[][]{1997ApJ...474L..43V}.

Following \citetalias{1997ApJ...474L..43V}, we estimated the coherence of \grs\ light curves obtained in different  energy bands. 
Because we use \nustar\ data (covering 3--78~keV energy band) we adopted the following energy bands for our analysis: 3--5, 5--8, 8--15 and 15--78~keV.
This partition of the \nustar\ energy band pursues the following idea: the energy spectrum of \grs\ can be described with two major components (see Section~\ref{sec:spec}): the power-law continuum and the reflection component consisting of the fluorescent Fe K$\alpha$ line and the Compton hump.
In the 5--8~keV band there is contribution of the prominent Fe K$\alpha$ line.
Taking into account its equivalent width of 0.2~keV it provides about 5 per cent of the flux in this band.
In the 8--15~keV energy band we expect mostly the power-law component to be present.
The Compton hump, another reflection feature, is stronger in the 15--78~keV energy band. 

As it was mentioned above the \nustar\ detectors have a complex dead-time depending on energy.
The coherence computed from one detector is subject to the dead-time cross-talk effects, i.e. capturing of the photon in a particular energy band prevents the registration of any next photon arriving during the dead-time \citep[see e.g.][]{2015MNRAS.451.4253R}. 
Such a cross-talk makes random independent processes more coherent.
In order to eliminate these effects in coherence estimation we follow the recipe suggested by \citet{2015ApJ...800..109B} for the cospectrum estimation. 
As explained in \citet{2015ApJ...800..109B} we can take advantage of the presence of two detectors modules, signals from which are processed independently. 
That means that the photon registered by one of the modules does not prevent registration of the photon arriving during the dead-time in another module. 
Therefore, for the nominator in equation~(\ref{eq:nowak_coh}) (cross product of the Fourier transforms of the light curves obtained in different energy bands) we can use light curves obtained from different modules, e.g., a light curve obtained in the soft band on the FPMA module with one obtained in the hard band on the FPMB module and vice versa.

To obtain proper estimation of the coherence it is also important to have the correct estimation of the intrinsic variability power spectrum (denominator in equation~(\ref{eq:nowak_coh})).
We use in this work a model independent approach, with the cospectrum used for the intrinsic variability power spectrum estimation (another approach would be to use Poisson-noise subtracted power spectrum of the original light curves or analytical function fitted to the power spectrum, obtained in the previous section).

The $P_{\rm unc}^2$ component was computed as suggested by \citetalias{1997ApJ...474L..43V}. 
The Poisson noise level was estimated as the mean power in the 5--15~Hz range.
In this frequency band the Poisson noise dominates over the source intrinsic variability, while its shape is not yet affected by the dead-time effects (the power spectrum is flat below 15~Hz).

There is one drawback in using the cospectrum for the estimation of intrinsic variability power spectrum. 
As was discussed in the previous section, the cospectrum can be described with Laplace statistics, which has non-zero probability density in the vicinity of zero and a positive mean value.
Therefore, if insufficient number of samples is used to calculate the mean of the cospectrum a large statistical error would be introduced to the coherence (because the cospectrum comes in the denominator, as it is used for the power spectrum estimation).  
The number of the samples is limited by the total duration of the observation and the condition that the shape of the cospectrum should not change significantly (otherwise an artificial dispersion would be introduced in the cospectrum distribution).
The last criterion appears to be the strictest one, because the QPO and the break frequencies change by a factor of two during the observation.
In order to increase the statistical significance of the estimated cospectrum we use the following property found to be inherent for XRBs intrinsic variability. 
\citet{wijnands99} showed that the primary features of the power spectrum of the XRBs in LHS and HIMS evolve simultaneously, i.e. the break frequency of the flat-top broad-band noise and the QPO centroid frequency are connected by the relation $f_{\rm b} \approx 0.3 f_{\rm QPO}$.
Taking into account this property of the power spectrum and the small scatter of the $f_{\rm b}/f_{\rm QPO}$ in our data, we stacked all 13 orbits, scaling the frequencies to preserve the QPO position. 
We assumed that the coherence in each tracked frequency channel is preserved along the observation and is scaled in a similar way.

\begin{figure}
    \includegraphics[width=\columnwidth]{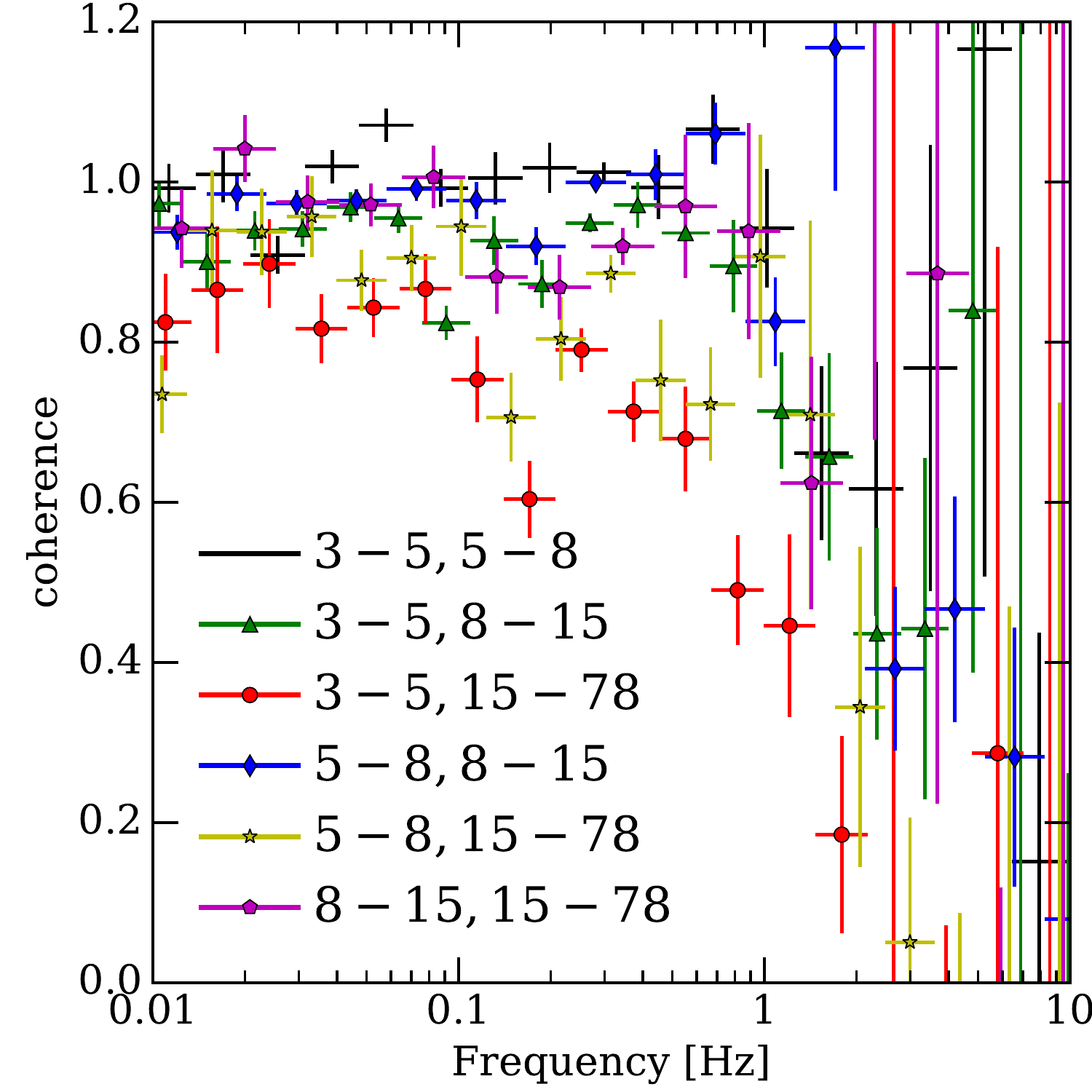}
    \caption{The coherence between light curves extracted in different pairs of energy bands: black dots, red circles, green triangles, blue diamonds, yellow asterisks and magenta pentagons are for (3--5, 5--8), (3--5, 8--15), (3--5, 15--78), (5--8, 8--15), (5--8, 15--78) and (8--15, 15--78)~keV, respectively.}
    \label{fig:coherence}
\end{figure}

The coherence between hard and soft energy bands at frequencies up to $\sim3$~Hz is presented in Fig.~\ref{fig:coherence}. 
We found that the coherence in the adjacent energy bands is close to unity, with the average values of $1.0\pm0.05$ in the 0.01--1~Hz frequency band.
However, for the 3--5 and 15--78~keV energy bands the coherence is significantly lower (Fig.~\ref{fig:coherence}). 
It is nearly constant $\approx0.85$ in the 0.005--0.1~Hz frequency band and drops down at higher frequencies.
Similar behaviour was observed for GX 339--4 \citepalias{1997ApJ...474L..43V}.

\begin{figure*}
\includegraphics[width=0.99\columnwidth]{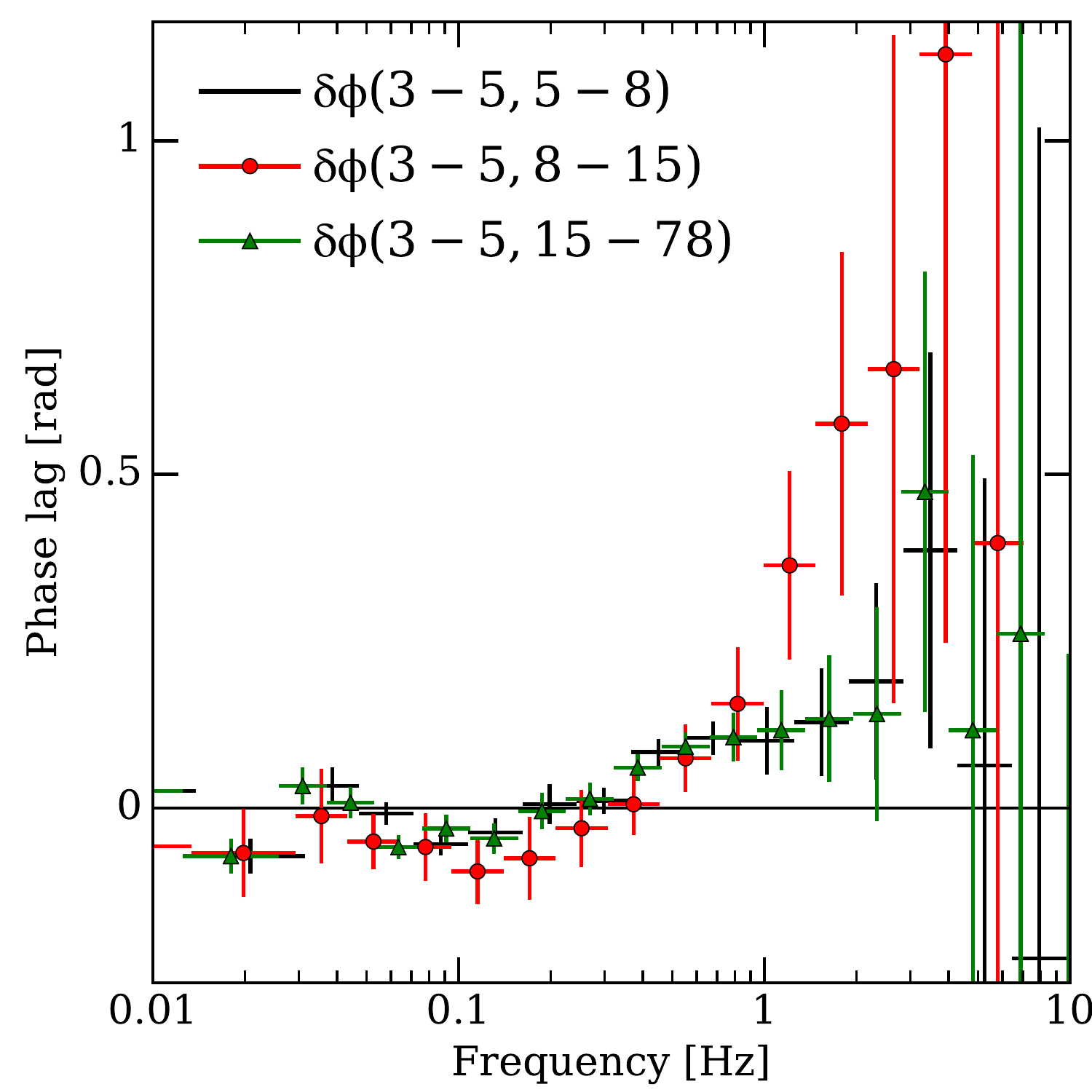}
\includegraphics[width=0.99\columnwidth]{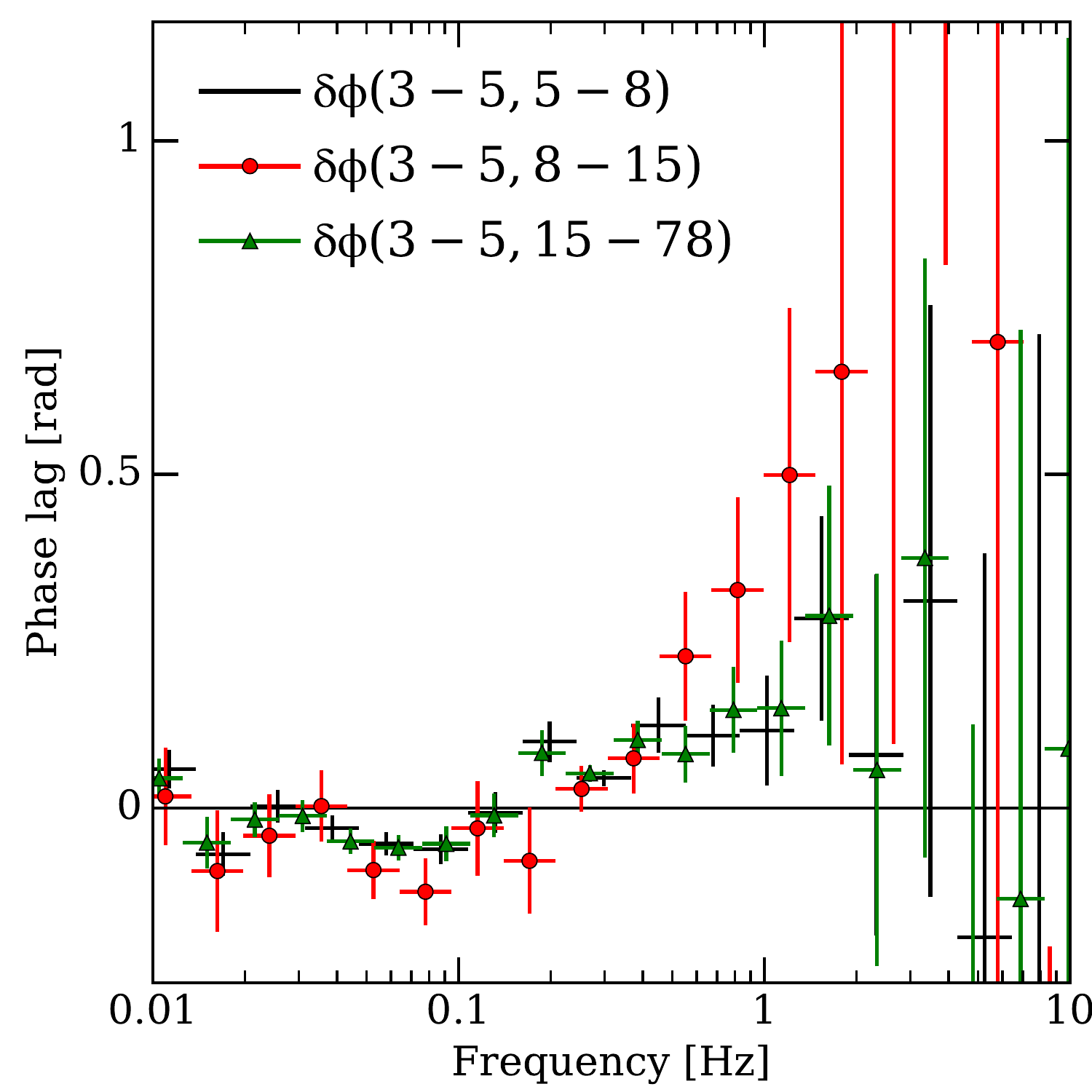}
\caption{Phase lags between the hard (5--8; 8--15; 15--78~keV) and soft (3--5~keV) energy bands. 
Left: The phase lag spectrum obtained from \nustar\ observations by stacking all data. Right: The phase lag spectrum obtained with tracked frequency, i.e. frequencies for each separate light-curve segment were scaled to conserve the QPO centroid at 0.3~Hz. }
\label{fig:phase_lag}
\end{figure*}

\subsection{Phase lags}

From the definition of the coherence (see equation (\ref{eq:nowak_coh})) it follows that the coherent signals have roughly constant phase shifts between their Fourier transforms in each frequency bin. 
To estimate the phase lags one has to calculate the average of the product of the Fourier transforms obtained in one energy band to the conjugated Fourier transforms estimated in another energy band.
The phase of the obtained complex value will be the  frequency-dependent phase lag
\begin{equation}
      \delta \phi(f) = \arctan{ \left( \frac{ \Im{\langle{F_{\rm s}^{*}(f) F_{\rm h}(f)}\rangle} }{\Re{\langle{F_{\rm s}^{*}(f) F_{\rm h}(f)}\rangle}} \right)},
\end{equation}
where $\Im$ and $\Re$ stand for the imaginary and real part of the complex value, respectively. 
For the uncertainty estimation we used the approach proposed by \citet{2014A&ARv..22...72U}: the phase lag uncertainty is caused by incoherent processes, therefore we take $\Delta\delta\phi \approx \arctan{(\Delta C(f)/C(f))}$, where $C(f)$ is the coherence and $\Delta C(f)$ is the coherence uncertainty.

The phase lags, observed for different systems, have features, which correlate with the power spectrum and those which have no obvious counterparts in it. 
Therefore, bearing in mind property of the linear evolution of all frequencies in the power spectra discussed in the previous section, we computed the phase lag spectrum with two approaches: with and without the tracing of the QPO centroid frequency  (Fig.\ref{fig:phase_lag}).
The obtained phase lags appears to be surprisingly similar, however, those calculated with the traced QPO frequency seem to have larger amplitude at lower frequencies, which may indicate that the phase lags indeed evolve in a similar way with the power spectrum. 
It appears that in the 0.5--3~Hz frequency band, the positive (hard) lags are present while at frequencies below 0.1~Hz there is an indication of the negative (soft) lag.
The observed phase lags correspond to the delay times between soft and hard photons $\sim$0.1~s for frequencies above 0.5 Hz and $-0.1$...$-1$~s for frequencies below 0.1 Hz.
Unfortunately, because of  an insufficient signal-to-noise ratio, we cannot determine whether there any specific features present at the QPO or its second harmonic centroid frequencies.

We also investigated the phase-lag energy dependence for two frequency bands where they are most prominent (0.5--1 and 1--5~Hz). 
We divided the \nustar\ energy band on 15 logarithmically spaced bins and computed the average phase lag between the light curve in the first energy bin (3--3.72~keV) and all the following bins.
Using the {\it RXTE} observatory data, \citet{2001MNRAS.327..799K} found logarithmic growth of the phase-lag with energy, however, the low signal-to-noise ratio of our data does not allow to robustly derive such a trend. 
At frequencies above 0.5~Hz the phase lag of the harder emission relative to the softest \nustar\  channels grows up to $\approx \pi/16$ at  energies above $\approx7$~keV and stays relatively constant above it.

\section{Discussion}
\label{sec:disc}

We have studied the spectro-timing evolution of \grs\, during its HIMS observed by \nustar\ during 2014 outburst. 
We performed spectral analysis, in order to determine the geometry of the accretion flow.
We also used different metrics of the system timing properties in order to additionally constrain its geometry and models describing the physical origin of the observed energy spectrum. 

\subsection{Disc}

With available quasi-simultaneous observation performed by \swiftx\ and \nustar\ we were able to obtain broad-band energy spectrum from 0.8 to 78~keV.
The spectrum contains strong reflection features - the broad emission line at energy $\sim6.5$~keV and the Compton-hump at $\approx 20$ keV. It is widely accepted, that such a spectrum can be produced by reflection of a primary emission, that originates from Comptonization of some seed soft photons in the hot corona or the inner hot flow, from the cold accretion disc. 
We appied \textsc{relxilllp} model in order to estimate some properties of the accretion flow, e.g. the inner radius of the cold disc.
Although the quality of the data prevented us from measuring the movement of the inner disc boundary throughout the observation,  the average broad-band energy spectrum implies the truncation radius smaller than 7.3 $GM/c^{2}$ (90 per cent confidence limit), which is in agreement with an estimate by  \citet{miller15_nust}. 

\subsection{Timing properties}

We also studied \grs\ by means of its power spectrum and the coherence function. 
Using \nustar\, data we found a prominent type-C LF QPO in its power spectrum, with the frequency monotonically growing along the observation. 
The power spectra are typical for the HIMS and consist of the broad-band noise and fundamental QPO, the second harmonic of the QPO is also seen. 

During several orbits from the first half of the observation, a subharmonic was also observed. 
In all 13 orbits, the second QPO harmonic is more prominent in the soft band (3--5 keV), with the ratio of its amplitude to that of the fundamental QPO being $0.565\pm0.02$ in the 3--5 keV band versus $0.275\pm0.02$ in the 15--78 keV band.
This indicates that the QPO pulse profile differs in two energy bands, similarly to the results of \citet{2015MNRAS.446.3516I}.
The amplitude of the fundamental QPO  correlates with the QPO frequency as well as with the spectral slope (see Fig.~\ref{fig:qpo_gamma}).
The measured velocity of the QPO frequency drift is found to be $\approx6.0\times10^{-6}$~Hz~s$^{-1}$.

The coherence measured between the adjacent energy bands in 0.01--1 Hz was found to be nearly unity, while the coherence measured between the softest (3--5~keV) and the hardest used energy bands (15--78~keV) was lower and showed significant drop above 0.1~Hz. 
\citetalias{1997ApJ...474L..43V} discussed what could lead to such a loss of the coherence between different energy bands. 
Two main possible mechanisms were a non-linear transfer function between soft and hard bands and contribution of several coherent but independent processes in each energy band.
Both these mechanisms can be related to the model of propagating fluctuations. 
In the first case, we can assume that the emission region is compact and the variability power spectrum is already formed in the outer parts of the flow, however, the fluxes in the soft and hard energy bands are connected non-linearly, thus coherence is lost. For example, let us assume that the photon flux is a power law $N(E)\propto E^{-\Gamma(t)}$, and its variability is caused by variation of $\Gamma$ (e.g. because of changes in the optical depth of the Comptonizing medium). 
Then the fluxes in the two energy bands around $E_1 $ and $E_2$ are connected by a power-law relation:
\begin{equation*}
N_2(t) \propto E_2^{-\Gamma(t)}  \propto \left(E_1^{- \Gamma(t)}\right)^\nu \propto N_1^\nu(t)
\end{equation*}
where $\nu = \log_{E_1}{E_2}$.
It then follows that the coherence loss will be increased with the energy bands separation.

In the second scenario, the emission zone is extended, with separate parts of the accretion flow being mainly responsible for the emission in the different energy bands.
When the accreted matter moves from the zone responsible for the soft (or hard) emission, having Fourier function $F_1(f)$, to the zone responsible for the emission in a different energy band, the initial variations are partially washed out because of the viscous propagation and new stochastic variability is injected $F_2(f) = F_1(f)G(f) + F_3(f)$. Where $G(f)$ is Green's function describing propagation of signals from zone 1 to zone 2. 
We would assume that the variability injected between zones 2 and 1 is independent to that observed in zone 1 : $|\langle F_1^*(f)F_3(f) \rangle|^2 = 0$.
It follows then that the coherence 
\begin{equation*}
C(f) = \frac{|\langle F_1(f)^*[G(f)F_1(f) + F_3(f)]\rangle|^2}{\langle|G(f)F_1(f) + F_3(f)|^2\rangle \langle|F_1(f)|^2\rangle} \approx 1 - \frac{\langle|F_3(f)|^2\rangle}{\langle|F_1(f)|^2\rangle}
\end{equation*}
will decrease as a result of evolution  of the power spectrum within the accretion flow.

\subsection{Phase lag}

The phase lags in different BH system were investigated by many authors \citep[see, e.g.][]{2003A&A...407..335M, 2006A&A...449..703R, 2011A&A...533A...8B, 2011MNRAS.415..292M, 2013MNRAS.435.2132M, 2017MNRAS.471.1475D}.
It was found that for stellar mass BHs, the phase lag in the frequency range occupied by the flat-top noise and LF QPOs is usually hard.
The time lags can be described with the power law $\delta \tau=\delta \phi /(2\pi f) \propto f^{-0.7}$ \citep{1989Natur.342..773M, 1999ApJ...517..355N} with positive or negative peaks at the frequencies of the QPO and its harmonics \citep{2000ApJ...531L..45C,Remillard02, 2005ApJ...622..508K, 2015MNRAS.449.4027A}. 
\citet{1989Natur.342..773M} tried to explain the observed lags with the clumpy flow model, which previously was used to explain the observed shape of the flat-top noise Fourier spectrum.
\citet{1999MNRAS.306L..31P} argued that the observed hard lags (and also the shape of the power spectrum) can be explained with the magnetic flares model. 
In that model the variability was caused by magnetic flares, while their spectral evolution produces the lags. 
Furthermore, X-ray reverberation may produce additional lags at energies where Compton reflection contributes \citep{2002MNRAS.332..257P}. 
To explain the observed hard lags and their dependence on the frequency, \citet{1999ApJ...515..726N} considered two models:  Comptonization in the extended corona \citep{1997ApJ...480..735K} and propagation of the perturbations in the advective flow \citep{1997MNRAS.292..679L}.  
The authors found that it is hard to explain the observed lags with either models, as the first one demanded a very extended corona ($\sim150 R_{\rm g}$) and the second one required a very slow propagation speed in the flow.
Later \citet{2001MNRAS.327..799K} reconsidered their result and, on the basis on the amplitude and the energy dependence of the hard lags, derived that they cannot be caused by the reverberation and are likely produced by propagation of the perturbation in the corona on the viscous time scale \citep[see also][for simulations]{2006MNRAS.367..801A}.
It is worth mentioning, that most of the proposed models generally can explain hard lags but fail to explain soft lags, which were later found in many sources both at low and high frequencies, below and above flat-top noise break frequency \citep{2000ApJ...531L..45C, 
2012MNRAS.427.2985C, 2017MNRAS.465.1926Y, 2017MNRAS.464.2643V}. 
Recently, \citet{2018MNRAS.474.2259M} showed that the soft lags are possible in the propagating fluctuations model if the outward movement of the disc surface density perturbations due to the viscous evolution are also considered.  

\citet{2017ApJ...845..143Z} showed that in the BH GX~339--4 the phase lags at the QPO frequency and its second harmonic evolve with the QPO frequency. 
\citet{2018MNRAS.473.4644R} found that the mean time lag strongly correlates with the photon index of power-law continuum, with the time lags increasing with decreasing hardness. 
They proposed, that the observed behaviour can be explained with the Comptonization of soft photons by energetic electrons in a jet.
\citet{2017MNRAS.464.2643V} found that the sign and the amplitude of the phase lag at the QPO frequency depend on the system inclination.

In the present case, we found that the time lags between hard (15--78~keV) and soft (3--5~keV) energy bands is $\approx+0.1$~s (hard) in the 0.1--3~Hz frequency range and $-1$...$-0.1$~s (soft) below 0.1~Hz.
The phase lag grows roughly linearly in the 0.1--2~Hz frequency range (i.e. the time delay between hard and soft emission is roughly constant).
In the frame of the propagating fluctuations model it can be explained if zones responsible for the soft and hard emission are separated by $R \approx \left[\left(H/R\right)^2 \tau \alpha \sqrt{2GM}\right]^{2/3} \approx 20 R_{\rm g}$ (for a 10M$_\odot$ BH with viscosity parameter $\alpha\approx 0.1$ and the height-to-radius ratio $H/R\approx 1$).

The obtained phase lags cannot be used to estimate the system inclination, with the dependence found by \citet{2017MNRAS.464.2643V}, because this dependence is valid only for the type-C QPO above 3 Hz.
We found very small dependence of the phase lags with energy, with the lags growing up to $\sim5$~keV and staying relatively constant above that energy.

\subsection{Correlation between spectral and timing properties}

During the \nustar\ observation both the energy spectrum and the power spectrum have gradually evolved. 
As the QPO frequency increases from 0.3 to 0.7~Hz the energy spectrum became softer: the power-law index grew from 1.47 to 1.55 and the cut-off energy decreased from 31 to 27 keV. Along with change in QPO frequency its amplitude also increased.
In the framework of the Lense-Thirring precession model for the QPO origin, increase of QPO frequency would imply that the inner edge of the cold accretion disc shrunk. The evolution of the energy spectrum with its general softening and lowering of the exponential cut-off temperature may also suggest the increasing role of the soft seed photons from the cold disc, which would be natural for the HIMS when the cold accretion disc moves closer to the BH and, correspondingly, to the region where Comptonization occurs.

\begin{figure}
\includegraphics[width=\columnwidth]{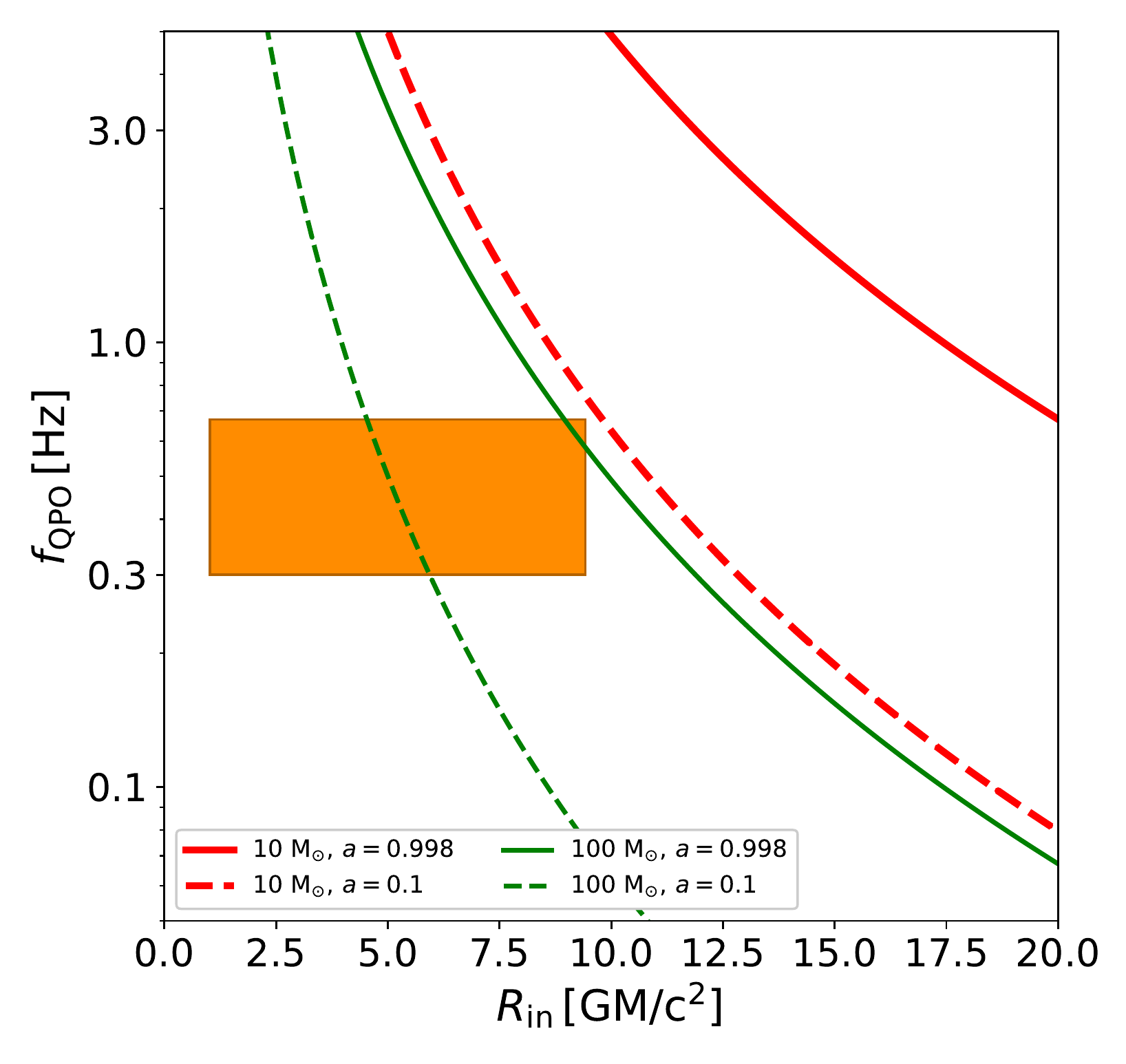}
\caption{Expected QPO frequency for a BH of a given mass and spin as a function of the disc inner radius  \citep{ingram14}. 
The orange square represents the region bounded by the observed QPO frequencies and the disc inner radius  measured from the spectra (99\% confidence interval). 
        }
        \label{fig:qpoconstr}
\end{figure}

\subsection{Black hole mass estimation}

If the type-C QPOs are caused by the Lense-Thirring precession of the inner part of the accretion disc  \citep{ingram09}, their frequencies should depend on the truncation radius.
We used the combination of the estimated disc inner radius and the observed QPO frequencies in order to assess the BH mass.
Following \citet{ingram14} we calculated the nodal precession frequency (which is thought to correspond to the QPO fundamental frequency) as a function of the disc inner radius for two values of the BH mass, 10 and 100$M_{\odot}$, and two values of the BH spin  $a=0.1$ and 0.998 (maximally rotating). 
We can see from Fig.~\ref{fig:qpoconstr} that observations are incompatible with the BH mass 10M$_{\odot}$ and barely agree with a maximally rotating massive 100M$_{\odot}$ BH. 
This results together with the measurements by \citet{fuerst16_gx339} and \citet{mereminskiy18_maxi} indicate that there are some tensions between the predictions of the Lense-Thirring precession model and the truncation radii inferred from the spectral fitting. 
It is also worth noting that the observed QPO changed in frequency by a factor of $\approx2.5$ while no drastic changes were observed in the spectrum.

\section{Summary} 

The discovery of the low-frequency QPO with rapidly changing frequency in the power spectra of the accreting black hole \grs\ during its 2014 outburst allowed us to probe connection between the timing and spectral properties of the X-ray emission in hard-intermediate state using \nustar\ and \swiftx\ data. 
While the analysis of the average energy spectrum points on the unusually small truncation radius ($R_{\rm in}<7.3\, R_{\rm g}$, see also \citealt{miller15_nust}) of the accretion disc, no signs of the disc black-body emission were found. 
The QPO frequency changes by a factor of 2.2 during the observation but no dramatic changes of the profile or equivalent width of the broadened iron line is seen, although the continuum emission softens as the QPO frequency grows. 

This implies that either the QPO frequency depends weakly on the truncation radius or the used spectral model underestimates the truncation radius. 
In the framework of the Lense-Thirring precession model, which associates the LF QPOs with the solid body precession of the whole inner part of the hot flow, a massive black hole with $M_{\rm BH} \approx 100$M$_\odot$ is required in order to simultaneously satisfy the constraints from the observed QPO frequency and the accretion disc truncation radius, inferred from the average spectrum.

\section*{Acknowledgements}

This work was supported by the Russian Science Foundation grant no. 14-12-01287. 
We thank E.~M.~Churazov for fruitful discussions and useful suggestions. 
We are grateful for T.~Dauser and J.~Gracia for their help with \textsc{relxill} model. 
This research has made use of data obtained through the High Energy Astrophysics Science Archive Research Center Online Service, provided by the NASA/Goddard Space Flight Center, as well as of data supplied by the UK Swift Science Data Centre at the University of Leicester.

\bibliographystyle{mnras}
\bibliography{author_en.bib,coherence.bib}

\begin{thebibliography}{}
\makeatletter
\relax
\def\mn@urlcharsother{\let\do\@makeother \do\$\do\&\do\#\do\^\do\_\do\%\do\~}
\def\mn@doi{\begingroup\mn@urlcharsother \@ifnextchar [ {\mn@doi@}
  {\mn@doi@[]}}
\def\mn@doi@[#1]#2{\def\@tempa{#1}\ifx\@tempa\@empty \href
  {http://dx.doi.org/#2} {doi:#2}\else \href {http://dx.doi.org/#2} {#1}\fi
  \endgroup}
\def\mn@eprint#1#2{\mn@eprint@#1:#2::\@nil}
\def\mn@eprint@arXiv#1{\href {http://arxiv.org/abs/#1} {{\tt arXiv:#1}}}
\def\mn@eprint@dblp#1{\href {http://dblp.uni-trier.de/rec/bibtex/#1.xml}
  {dblp:#1}}
\def\mn@eprint@#1:#2:#3:#4\@nil{\def\@tempa {#1}\def\@tempb {#2}\def\@tempc
  {#3}\ifx \@tempc \@empty \let \@tempc \@tempb \let \@tempb \@tempa \fi \ifx
  \@tempb \@empty \def\@tempb {arXiv}\fi \@ifundefined
  {mn@eprint@\@tempb}{\@tempb:\@tempc}{\expandafter \expandafter \csname
  mn@eprint@\@tempb\endcsname \expandafter{\@tempc}}}

\bibitem[\protect\citeauthoryear{{Allen}}{{Allen}}{1973}]{1973asqu.book.....A}
{Allen} C.~W.,  1973, {Astrophysical quantities}.
University of London, Athlone Press, London

\bibitem[\protect\citeauthoryear{{Altamirano} \& {M{\'e}ndez}}{{Altamirano} \&
  {M{\'e}ndez}}{2015}]{2015MNRAS.449.4027A}
{Altamirano} D.,  {M{\'e}ndez} M.,  2015, \mn@doi [\mnras]
  {10.1093/mnras/stv556}, \href
  {http://adsabs.harvard.edu/abs/2015MNRAS.449.4027A} {449, 4027}

\bibitem[\protect\citeauthoryear{{Ar{\'e}valo} \& {Uttley}}{{Ar{\'e}valo} \&
  {Uttley}}{2006}]{2006MNRAS.367..801A}
{Ar{\'e}valo} P.,  {Uttley} P.,  2006, \mn@doi [\mnras]
  {10.1111/j.1365-2966.2006.09989.x}, \href
  {http://adsabs.harvard.edu/abs/2006MNRAS.367..801A} {367, 801}

\bibitem[\protect\citeauthoryear{{Arnaud}}{{Arnaud}}{1996}]{arnaud96}
{Arnaud} K.~A.,  1996, in {Jacoby} G.~H.,  {Barnes} J.,  eds,  Astronomical
  Society of the Pacific Conference Series Vol. 101, Astronomical Data Analysis
  Software and Systems V. p.~17

\bibitem[\protect\citeauthoryear{{Bachetti} et~al.,}{{Bachetti}
  et~al.}{2015}]{2015ApJ...800..109B}
{Bachetti} M.,  et~al., 2015, \mn@doi [\apj] {10.1088/0004-637X/800/2/109},
  \href {http://adsabs.harvard.edu/abs/2015ApJ...800..109B} {800, 109}

\bibitem[\protect\citeauthoryear{{Belloni}}{{Belloni}}{2010}]{belloni10}
{Belloni} T.~M.,  2010, in {Belloni} T.,  ed.,  Lecture Notes in Physics Vol.
  794, The Jet Paradigm. Springer Verlag, Berlin, p.~53 (\mn@eprint {arXiv}
  {0909.2474}), \mn@doi{10.1007/978-3-540-76937-8_3}

\bibitem[\protect\citeauthoryear{{Belloni} \& {Hasinger}}{{Belloni} \&
  {Hasinger}}{1990}]{1990A&A...227L..33B}
{Belloni} T.,  {Hasinger} G.,  1990, \aap, \href
  {http://adsabs.harvard.edu/abs/1990A%26A...227L..33B} {227, L33}

\bibitem[\protect\citeauthoryear{{B{\"o}ck} et~al.,}{{B{\"o}ck}
  et~al.}{2011}]{2011A&A...533A...8B}
{B{\"o}ck} M.,  et~al., 2011, \mn@doi [\aap] {10.1051/0004-6361/201117159},
  \href {http://adsabs.harvard.edu/abs/2011A%26A...533A...8B} {533, A8}

\bibitem[\protect\citeauthoryear{{Borozdin} \& {Trudolyubov}}{{Borozdin} \&
  {Trudolyubov}}{2000}]{borozdin00}
{Borozdin} K.~N.,  {Trudolyubov} S.~P.,  2000, \mn@doi [\apjl]
  {10.1086/312625}, \href {http://adsabs.harvard.edu/abs/2000ApJ...533L.131B}
  {533, L131}

\bibitem[\protect\citeauthoryear{{Borozdin}, {Revnivtsev}, {Trudolyubov},
  {Aleksandrovich}, {Sunyaev}  \& {Skinner}}{{Borozdin}
  et~al.}{1998}]{borozdin98}
{Borozdin} K.~N.,  {Revnivtsev} M.~G.,  {Trudolyubov} S.~P.,  {Aleksandrovich}
  N.~L.,  {Sunyaev} R.~A.,   {Skinner} G.~K.,  1998, Astron. Lett., \href
  {http://adsabs.harvard.edu/abs/1998AstL...24..435B} {24, 435}

\bibitem[\protect\citeauthoryear{{Cabanac}, {Henri}, {Petrucci}, {Malzac},
  {Ferreira}  \& {Belloni}}{{Cabanac} et~al.}{2010}]{cabanac10}
{Cabanac} C.,  {Henri} G.,  {Petrucci} P.-O.,  {Malzac} J.,  {Ferreira} J.,
  {Belloni} T.~M.,  2010, \mn@doi [\mnras] {10.1111/j.1365-2966.2010.16340.x},
  \href {http://adsabs.harvard.edu/abs/2010MNRAS.404..738C} {404, 738}

\bibitem[\protect\citeauthoryear{{Casella}, {Belloni}  \& {Stella}}{{Casella}
  et~al.}{2005}]{casella05}
{Casella} P.,  {Belloni} T.,   {Stella} L.,  2005, \mn@doi [\apj]
  {10.1086/431174}, \href {http://adsabs.harvard.edu/abs/2005ApJ...629..403C}
  {629, 403}

\bibitem[\protect\citeauthoryear{{Cassatella}, {Uttley}  \&
  {Maccarone}}{{Cassatella} et~al.}{2012}]{2012MNRAS.427.2985C}
{Cassatella} P.,  {Uttley} P.,   {Maccarone} T.~J.,  2012, \mn@doi [\mnras]
  {10.1111/j.1365-2966.2012.22021.x}, \href
  {http://adsabs.harvard.edu/abs/2012MNRAS.427.2985C} {427, 2985}

\bibitem[\protect\citeauthoryear{{Churazov}, {Gilfanov}  \&
  {Revnivtsev}}{{Churazov} et~al.}{2001}]{2001MNRAS.321..759C}
{Churazov} E.,  {Gilfanov} M.,   {Revnivtsev} M.,  2001, \mn@doi [\mnras]
  {10.1046/j.1365-8711.2001.04056.x}, \href
  {http://adsabs.harvard.edu/abs/2001MNRAS.321..759C} {321, 759}

\bibitem[\protect\citeauthoryear{{Cui}, {Zhang}  \& {Chen}}{{Cui}
  et~al.}{2000}]{2000ApJ...531L..45C}
{Cui} W.,  {Zhang} S.~N.,   {Chen} W.,  2000, \mn@doi [\apjl] {10.1086/312520},
  \href {http://adsabs.harvard.edu/abs/2000ApJ...531L..45C} {531, L45}

\bibitem[\protect\citeauthoryear{{Dauser}, {Garc{\'{\i}}a}, {Parker}, {Fabian}
  \& {Wilms}}{{Dauser} et~al.}{2014}]{dauser14}
{Dauser} T.,  {Garc{\'{\i}}a} J.,  {Parker} M.~L.,  {Fabian} A.~C.,   {Wilms}
  J.,  2014, \mn@doi [\mnras] {10.1093/mnrasl/slu125}, \href
  {http://adsabs.harvard.edu/abs/2014MNRAS.444L.100D} {444, L100}

\bibitem[\protect\citeauthoryear{{Dauser}, {Garc{\'{\i}}a}, {Walton},
  {Eikmann}, {Kallman}, {McClintock}  \& {Wilms}}{{Dauser}
  et~al.}{2016}]{dauser16}
{Dauser} T.,  {Garc{\'{\i}}a} J.,  {Walton} D.~J.,  {Eikmann} W.,  {Kallman}
  T.,  {McClintock} J.,   {Wilms} J.,  2016, \mn@doi [\aap]
  {10.1051/0004-6361/201628135}, \href
  {http://adsabs.harvard.edu/abs/2016A%26A...590A..76D} {590, A76}

\bibitem[\protect\citeauthoryear{{De Marco} et~al.,}{{De Marco}
  et~al.}{2017}]{2017MNRAS.471.1475D}
{De Marco} B.,  et~al., 2017, \mn@doi [\mnras] {10.1093/mnras/stx1649}, \href
  {http://adsabs.harvard.edu/abs/2017MNRAS.471.1475D} {471, 1475}

\bibitem[\protect\citeauthoryear{{Dickey} \& {Lockman}}{{Dickey} \&
  {Lockman}}{1990}]{1990ARA&A..28..215D}
{Dickey} J.~M.,  {Lockman} F.~J.,  1990, \mn@doi [\araa]
  {10.1146/annurev.aa.28.090190.001243}, \href
  {http://adsabs.harvard.edu/abs/1990ARA%26A..28..215D} {28, 215}

\bibitem[\protect\citeauthoryear{{Done}, {Gierli{\'n}ski}  \& {Kubota}}{{Done}
  et~al.}{2007}]{DGK07}
{Done} C.,  {Gierli{\'n}ski} M.,   {Kubota} A.,  2007, \mn@doi [\aapr]
  {10.1007/s00159-007-0006-1}, \href
  {http://adsabs.harvard.edu/abs/2007A%26ARv..15....1D} {15, 1}

\bibitem[\protect\citeauthoryear{{Dunn}, {Fender}, {K{\"o}rding}, {Belloni}  \&
  {Cabanac}}{{Dunn} et~al.}{2010}]{2010MNRAS.403...61D}
{Dunn} R.~J.~H.,  {Fender} R.~P.,  {K{\"o}rding} E.~G.,  {Belloni} T.,
  {Cabanac} C.,  2010, \mn@doi [\mnras] {10.1111/j.1365-2966.2010.16114.x},
  \href {http://adsabs.harvard.edu/abs/2010MNRAS.403...61D} {403, 61}

\bibitem[\protect\citeauthoryear{{Eardley}, {Lightman}  \& {Shapiro}}{{Eardley}
  et~al.}{1975}]{1975ApJ...199L.153E}
{Eardley} D.~M.,  {Lightman} A.~P.,   {Shapiro} S.~L.,  1975, \mn@doi [\apjl]
  {10.1086/181871}, \href {http://adsabs.harvard.edu/abs/1975ApJ...199L.153E}
  {199, L153}

\bibitem[\protect\citeauthoryear{{Esin}, {McClintock}  \& {Narayan}}{{Esin}
  et~al.}{1997}]{Esin97}
{Esin} A.~A.,  {McClintock} J.~E.,   {Narayan} R.,  1997, \mn@doi [\apj]
  {10.1086/304829}, \href {http://adsabs.harvard.edu/abs/1997ApJ...489..865E}
  {489, 865}

\bibitem[\protect\citeauthoryear{{Evans} et~al.,}{{Evans}
  et~al.}{2009}]{evans09}
{Evans} P.~A.,  et~al., 2009, \mn@doi [\mnras]
  {10.1111/j.1365-2966.2009.14913.x}, \href
  {http://adsabs.harvard.edu/abs/2009MNRAS.397.1177E} {397, 1177}

\bibitem[\protect\citeauthoryear{{Filippova} et~al.,}{{Filippova}
  et~al.}{2014}]{filippova14}
{Filippova} E.,  et~al., 2014, The Astronomer's Telegram, \href
  {http://adsabs.harvard.edu/abs/2014ATel.5991....1F} {5991}

\bibitem[\protect\citeauthoryear{{Fragile}, {Blaes}, {Anninos}  \&
  {Salmonson}}{{Fragile} et~al.}{2007}]{FB07}
{Fragile} P.~C.,  {Blaes} O.~M.,  {Anninos} P.,   {Salmonson} J.~D.,  2007,
  \mn@doi [\apj] {10.1086/521092}, \href
  {http://adsabs.harvard.edu/abs/2007ApJ...668..417F} {668, 417}

\bibitem[\protect\citeauthoryear{{F{\"u}rst} et~al.,}{{F{\"u}rst}
  et~al.}{2016a}]{fuerst16_gx339}
{F{\"u}rst} F.,  et~al., 2016a, \mn@doi [\apj] {10.3847/0004-637X/828/1/34},
  \href {http://adsabs.harvard.edu/abs/2016ApJ...828...34F} {828, 34}

\bibitem[\protect\citeauthoryear{{F{\"u}rst} et~al.,}{{F{\"u}rst}
  et~al.}{2016b}]{fuerst16}
{F{\"u}rst} F.,  et~al., 2016b, \mn@doi [\apj] {10.3847/0004-637X/832/2/115},
  \href {http://adsabs.harvard.edu/abs/2016ApJ...832..115F} {832, 115}

\bibitem[\protect\citeauthoryear{{Garc{\'{\i}}a}, {Dauser}, {Reynolds},
  {Kallman}, {McClintock}, {Wilms}  \& {Eikmann}}{{Garc{\'{\i}}a}
  et~al.}{2013}]{garcia13}
{Garc{\'{\i}}a} J.,  {Dauser} T.,  {Reynolds} C.~S.,  {Kallman} T.~R.,
  {McClintock} J.~E.,  {Wilms} J.,   {Eikmann} W.,  2013, \mn@doi [\apj]
  {10.1088/0004-637X/768/2/146}, \href
  {http://adsabs.harvard.edu/abs/2013ApJ...768..146G} {768, 146}

\bibitem[\protect\citeauthoryear{Garc{\'{\i}}a et~al.,}{Garc{\'{\i}}a
  et~al.}{2014}]{garcia14}
Garc{\'{\i}}a J.,  et~al., 2014, \mn@doi [\apj] {10.1088/0004-637X/782/2/76},
  \href {http://cdsads.u-strasbg.fr/abs/2014ApJ...782...76G} {782, 76}

\bibitem[\protect\citeauthoryear{{Gierli{\'n}ski}, {Zdziarski}, {Poutanen},
  {Coppi}, {Ebisawa}  \& {Johnson}}{{Gierli{\'n}ski}
  et~al.}{1999}]{gierlinski99}
{Gierli{\'n}ski} M.,  {Zdziarski} A.~A.,  {Poutanen} J.,  {Coppi} P.~S.,
  {Ebisawa} K.,   {Johnson} W.~N.,  1999, \mn@doi [\mnras]
  {10.1046/j.1365-8711.1999.02875.x}, \href
  {http://adsabs.harvard.edu/abs/1999MNRAS.309..496G} {309, 496}

\bibitem[\protect\citeauthoryear{{Gierli{\'n}ski}, {Middleton}, {Ward}  \&
  {Done}}{{Gierli{\'n}ski} et~al.}{2008}]{gierlinski08}
{Gierli{\'n}ski} M.,  {Middleton} M.,  {Ward} M.,   {Done} C.,  2008, \mn@doi
  [\nat] {10.1038/nature07277}, \href
  {http://adsabs.harvard.edu/abs/2008Natur.455..369G} {455, 369}

\bibitem[\protect\citeauthoryear{{Gilfanov}}{{Gilfanov}}{2010}]{Gilfanov10}
{Gilfanov} M.,  2010, in {Belloni} T.,  ed.,  Lecture Notes in Physics, Berlin
  Springer Verlag Vol. 794, The Jet Paradigm. Springer Verlag, Berlin, p.~17
  (\mn@eprint {arXiv} {0909.2567}), \mn@doi{10.1007/978-3-540-76937-8_2}

\bibitem[\protect\citeauthoryear{{Gilfanov}, {Churazov}  \&
  {Revnivtsev}}{{Gilfanov} et~al.}{1999}]{GCR99}
{Gilfanov} M.,  {Churazov} E.,   {Revnivtsev} M.,  1999, \aap, \href
  {http://adsabs.harvard.edu/abs/1999A%26A...352..182G} {352, 182}

\bibitem[\protect\citeauthoryear{{Gilfanov}, {Churazov}  \&
  {Revnivtsev}}{{Gilfanov} et~al.}{2000}]{2000MNRAS.316..923G}
{Gilfanov} M.,  {Churazov} E.,   {Revnivtsev} M.,  2000, \mn@doi [\mnras]
  {10.1046/j.1365-8711.2000.03686.x}, \href
  {http://adsabs.harvard.edu/abs/2000MNRAS.316..923G} {316, 923}

\bibitem[\protect\citeauthoryear{{Goodman} \& {Weare}}{{Goodman} \&
  {Weare}}{2010}]{goodman10_gw}
{Goodman} J.,  {Weare} J.,  2010, \mn@doi [Communications in Applied
  Mathematics and Computational Science] {10.2140/camcos.2010.5.65}, \href
  {http://adsabs.harvard.edu/abs/2010CAMCS...5...65G} {5, 65}

\bibitem[\protect\citeauthoryear{{Grebenev} et~al.,}{{Grebenev}
  et~al.}{1993}]{grebenev93}
{Grebenev} S.,  et~al., 1993, \aaps, \href
  {http://adsabs.harvard.edu/abs/1993A%26AS...97..281G} {97, 281}

\bibitem[\protect\citeauthoryear{{Grebenev}, {Sunyaev}  \&
  {Pavlinsky}}{{Grebenev} et~al.}{1997}]{grebenev97}
{Grebenev} S.~A.,  {Sunyaev} R.~A.,   {Pavlinsky} M.~N.,  1997, \mn@doi [Adv.
  Space Res.] {10.1016/S0273-1177(97)00031-8}, \href
  {http://adsabs.harvard.edu/abs/1997AdSpR..19...15G} {19, 15}

\bibitem[\protect\citeauthoryear{{Greiner}, {Dennerl}  \& {Predehl}}{{Greiner}
  et~al.}{1996}]{greiner96}
{Greiner} J.,  {Dennerl} K.,   {Predehl} P.,  1996, \aap, \href
  {http://adsabs.harvard.edu/abs/1996A%26A...314L..21G} {314, L21}

\bibitem[\protect\citeauthoryear{{Harrison} et~al.,}{{Harrison}
  et~al.}{2013}]{harrison13_nust}
{Harrison} F.~A.,  et~al., 2013, \mn@doi [\apj] {10.1088/0004-637X/770/2/103},
  \href {http://adsabs.harvard.edu/abs/2013ApJ...770..103H} {770, 103}

\bibitem[\protect\citeauthoryear{{Heil}, {Uttley}  \& {Klein-Wolt}}{{Heil}
  et~al.}{2015}]{heil15}
{Heil} L.~M.,  {Uttley} P.,   {Klein-Wolt} M.,  2015, \mn@doi [\mnras]
  {10.1093/mnras/stv191}, \href
  {http://adsabs.harvard.edu/abs/2015MNRAS.448.3339H} {448, 3339}

\bibitem[\protect\citeauthoryear{{Homan} \& {Belloni}}{{Homan} \&
  {Belloni}}{2005}]{homan05}
{Homan} J.,  {Belloni} T.,  2005, \mn@doi [\apss] {10.1007/s10509-005-1197-4},
  \href {http://adsabs.harvard.edu/abs/2005Ap%26SS.300..107H} {300, 107}

\bibitem[\protect\citeauthoryear{{Huppenkothen} \& {Bachetti}}{{Huppenkothen}
  \& {Bachetti}}{2018}]{2018ApJS..236...13H}
{Huppenkothen} D.,  {Bachetti} M.,  2018, \mn@doi [\apjs]
  {10.3847/1538-4365/aabe38}, \href
  {http://adsabs.harvard.edu/abs/2018ApJS..236...13H} {236, 13}

\bibitem[\protect\citeauthoryear{{Ibragimov}, {Poutanen}, {Gilfanov},
  {Zdziarski}  \& {Shrader}}{{Ibragimov} et~al.}{2005}]{Ibragimov05}
{Ibragimov} A.,  {Poutanen} J.,  {Gilfanov} M.,  {Zdziarski} A.~A.,   {Shrader}
  C.~R.,  2005, \mn@doi [\mnras] {10.1111/j.1365-2966.2005.09415.x}, \href
  {http://adsabs.harvard.edu/abs/2005MNRAS.362.1435I} {362, 1435}

\bibitem[\protect\citeauthoryear{{Ingram} \& {Motta}}{{Ingram} \&
  {Motta}}{2014}]{ingram14}
{Ingram} A.,  {Motta} S.,  2014, \mn@doi [\mnras] {10.1093/mnras/stu1585},
  \href {http://adsabs.harvard.edu/abs/2014MNRAS.444.2065I} {444, 2065}

\bibitem[\protect\citeauthoryear{{Ingram} \& {van der Klis}}{{Ingram} \& {van
  der Klis}}{2013}]{2013MNRAS.434.1476I}
{Ingram} A.,  {van der Klis} M.,  2013, \mn@doi [\mnras]
  {10.1093/mnras/stt1107}, \href
  {http://adsabs.harvard.edu/abs/2013MNRAS.434.1476I} {434, 1476}

\bibitem[\protect\citeauthoryear{{Ingram} \& {van der Klis}}{{Ingram} \& {van
  der Klis}}{2015}]{2015MNRAS.446.3516I}
{Ingram} A.,  {van der Klis} M.,  2015, \mn@doi [\mnras]
  {10.1093/mnras/stu2373}, \href
  {http://adsabs.harvard.edu/abs/2015MNRAS.446.3516I} {446, 3516}

\bibitem[\protect\citeauthoryear{{Ingram}, {Done}  \& {Fragile}}{{Ingram}
  et~al.}{2009}]{ingram09}
{Ingram} A.,  {Done} C.,   {Fragile} P.~C.,  2009, \mn@doi [\mnras]
  {10.1111/j.1745-3933.2009.00693.x}, \href
  {http://adsabs.harvard.edu/abs/2009MNRAS.397L.101I} {397, L101}

\bibitem[\protect\citeauthoryear{{Ji}, {Zhang}, {Qu}  \& {Li}}{{Ji}
  et~al.}{2003}]{2003ApJ...584L..23J}
{Ji} J.~F.,  {Zhang} S.~N.,  {Qu} J.~L.,   {Li} T.~P.,  2003, \mn@doi [\apjl]
  {10.1086/368269}, \href {http://adsabs.harvard.edu/abs/2003ApJ...584L..23J}
  {584, L23}

\bibitem[\protect\citeauthoryear{{Kalberla}, {Burton}, {Hartmann}, {Arnal},
  {Bajaja}, {Morras}  \& {P{\"o}ppel}}{{Kalberla}
  et~al.}{2005}]{2005A&A...440..775K}
{Kalberla} P.~M.~W.,  {Burton} W.~B.,  {Hartmann} D.,  {Arnal} E.~M.,  {Bajaja}
  E.,  {Morras} R.,   {P{\"o}ppel} W.~G.~L.,  2005, \mn@doi [\aap]
  {10.1051/0004-6361:20041864}, \href
  {http://adsabs.harvard.edu/abs/2005A%26A...440..775K} {440, 775}

\bibitem[\protect\citeauthoryear{{Kalemci}, {Tomsick}, {Buxton}, {Rothschild},
  {Pottschmidt}, {Corbel}, {Brocksopp}  \& {Kaaret}}{{Kalemci}
  et~al.}{2005}]{2005ApJ...622..508K}
{Kalemci} E.,  {Tomsick} J.~A.,  {Buxton} M.~M.,  {Rothschild} R.~E.,
  {Pottschmidt} K.,  {Corbel} S.,  {Brocksopp} C.,   {Kaaret} P.,  2005,
  \mn@doi [\apj] {10.1086/427818}, \href
  {http://adsabs.harvard.edu/abs/2005ApJ...622..508K} {622, 508}

\bibitem[\protect\citeauthoryear{{Kazanas}, {Hua}  \& {Titarchuk}}{{Kazanas}
  et~al.}{1997}]{1997ApJ...480..735K}
{Kazanas} D.,  {Hua} X.-M.,   {Titarchuk} L.,  1997, \mn@doi [\apj]
  {10.1086/303991}, \href {http://adsabs.harvard.edu/abs/1997ApJ...480..735K}
  {480, 735}

\bibitem[\protect\citeauthoryear{{Kotov}, {Churazov}  \& {Gilfanov}}{{Kotov}
  et~al.}{2001}]{2001MNRAS.327..799K}
{Kotov} O.,  {Churazov} E.,   {Gilfanov} M.,  2001, \mn@doi [\mnras]
  {10.1046/j.1365-8711.2001.04769.x}, \href
  {http://adsabs.harvard.edu/abs/2001MNRAS.327..799K} {327, 799}

\bibitem[\protect\citeauthoryear{{Krimm} et~al.,}{{Krimm}
  et~al.}{2013}]{krimm13bat}
{Krimm} H.~A.,  et~al., 2013, \mn@doi [\apjs] {10.1088/0067-0049/209/1/14},
  \href {http://adsabs.harvard.edu/abs/2013ApJS..209...14K} {209, 14}

\bibitem[\protect\citeauthoryear{{Krimm} et~al.,}{{Krimm}
  et~al.}{2014}]{krimm14_atel}
{Krimm} H.~A.,  et~al., 2014, The Astronomer's Telegram, \href
  {http://adsabs.harvard.edu/abs/2014ATel.5986....1K} {5986}

\bibitem[\protect\citeauthoryear{{Lyubarskii}}{{Lyubarskii}}{1997}]{1997MNRAS.292..679L}
{Lyubarskii} Y.~E.,  1997, \mn@doi [\mnras] {10.1093/mnras/292.3.679}, \href
  {http://adsabs.harvard.edu/abs/1997MNRAS.292..679L} {292, 679}

\bibitem[\protect\citeauthoryear{{Maccarone}, {Coppi}  \&
  {Poutanen}}{{Maccarone} et~al.}{2000}]{2000ApJ...537L.107M}
{Maccarone} T.~J.,  {Coppi} P.~S.,   {Poutanen} J.,  2000, \mn@doi [\apjl]
  {10.1086/312778}, \href {http://adsabs.harvard.edu/abs/2000ApJ...537L.107M}
  {537, L107}

\bibitem[\protect\citeauthoryear{{Malzac} \& {Belmont}}{{Malzac} \&
  {Belmont}}{2009}]{MB09}
{Malzac} J.,  {Belmont} R.,  2009, \mn@doi [\mnras]
  {10.1111/j.1365-2966.2008.14142.x}, \href
  {http://adsabs.harvard.edu/abs/2009MNRAS.392..570M} {392, 570}

\bibitem[\protect\citeauthoryear{{Malzac}, {Belloni}, {Spruit}  \&
  {Kanbach}}{{Malzac} et~al.}{2003}]{2003A&A...407..335M}
{Malzac} J.,  {Belloni} T.,  {Spruit} H.~C.,   {Kanbach} G.,  2003, \mn@doi
  [\aap] {10.1051/0004-6361:20030859}, \href
  {http://adsabs.harvard.edu/abs/2003A%26A...407..335M} {407, 335}

\bibitem[\protect\citeauthoryear{{Marshall}, {Robin}, {Reyl{\'e}}, {Schultheis}
   \& {Picaud}}{{Marshall} et~al.}{2006}]{2006A&A...453..635M}
{Marshall} D.~J.,  {Robin} A.~C.,  {Reyl{\'e}} C.,  {Schultheis} M.,   {Picaud}
  S.,  2006, \mn@doi [\aap] {10.1051/0004-6361:20053842}, \href
  {http://adsabs.harvard.edu/abs/2006A%26A...453..635M} {453, 635}

\bibitem[\protect\citeauthoryear{{Matsuoka} et~al.,}{{Matsuoka}
  et~al.}{2009}]{matsuoka13maxi}
{Matsuoka} M.,  et~al., 2009, \mn@doi [\pasj] {10.1093/pasj/61.5.999}, \href
  {http://ads.nao.ac.jp/abs/2009PASJ...61..999M} {61, 999}

\bibitem[\protect\citeauthoryear{{Mauche}}{{Mauche}}{2002}]{mauche02}
{Mauche} C.~W.,  2002, \mn@doi [\apj] {10.1086/343095}, \href
  {http://adsabs.harvard.edu/abs/2002ApJ...580..423M} {580, 423}

\bibitem[\protect\citeauthoryear{{M{\'e}ndez}, {Altamirano}, {Belloni}  \&
  {Sanna}}{{M{\'e}ndez} et~al.}{2013}]{2013MNRAS.435.2132M}
{M{\'e}ndez} M.,  {Altamirano} D.,  {Belloni} T.,   {Sanna} A.,  2013, \mn@doi
  [\mnras] {10.1093/mnras/stt1431}, \href
  {http://adsabs.harvard.edu/abs/2013MNRAS.435.2132M} {435, 2132}

\bibitem[\protect\citeauthoryear{{Mereminskiy}, {Filippova}, {Krivonos},
  {Grebenev}, {Burenin}  \& {Sunyaev}}{{Mereminskiy}
  et~al.}{2017}]{mereminskiy17grs}
{Mereminskiy} I.~A.,  {Filippova} E.~V.,  {Krivonos} R.~A.,  {Grebenev} S.~A.,
  {Burenin} R.~A.,   {Sunyaev} R.~A.,  2017, \mn@doi [Astronomy Letters]
  {10.1134/S1063773717030057}, \href
  {http://adsabs.harvard.edu/abs/2017AstL...43..167M} {43, 167}

\bibitem[\protect\citeauthoryear{{Mereminskiy}, {Grebenev}, {Prosvetov}  \&
  {Semena}}{{Mereminskiy} et~al.}{2018}]{mereminskiy18_maxi}
{Mereminskiy} I.~A.,  {Grebenev} S.~A.,  {Prosvetov} A.~V.,   {Semena} A.~N.,
  2018, \mn@doi [Astronomy Letters] {10.1134/S106377371806004X}, \href
  {http://adsabs.harvard.edu/abs/2018AstL...44..378M} {44, 378}

\bibitem[\protect\citeauthoryear{{Miller}, {Homan}  \& {Miniutti}}{{Miller}
  et~al.}{2006a}]{miller06b}
{Miller} J.~M.,  {Homan} J.,   {Miniutti} G.,  2006a, \mn@doi [\apjl]
  {10.1086/510015}, \href {http://adsabs.harvard.edu/abs/2006ApJ...652L.113M}
  {652, L113}

\bibitem[\protect\citeauthoryear{{Miller}, {Homan}, {Steeghs}, {Rupen},
  {Hunstead}, {Wijnands}, {Charles}  \& {Fabian}}{{Miller}
  et~al.}{2006b}]{miller06a}
{Miller} J.~M.,  {Homan} J.,  {Steeghs} D.,  {Rupen} M.,  {Hunstead} R.~W.,
  {Wijnands} R.,  {Charles} P.~A.,   {Fabian} A.~C.,  2006b, \mn@doi [\apj]
  {10.1086/508644}, \href {http://adsabs.harvard.edu/abs/2006ApJ...653..525M}
  {653, 525}

\bibitem[\protect\citeauthoryear{{Miller} et~al.,}{{Miller}
  et~al.}{2015}]{miller15_nust}
{Miller} J.~M.,  et~al., 2015, \mn@doi [\apjl] {10.1088/2041-8205/799/1/L6},
  \href {http://adsabs.harvard.edu/abs/2015ApJ...799L...6M} {799, L6}

\bibitem[\protect\citeauthoryear{{Miyamoto} \& {Kitamoto}}{{Miyamoto} \&
  {Kitamoto}}{1989}]{1989Natur.342..773M}
{Miyamoto} S.,  {Kitamoto} S.,  1989, \mn@doi [\nat] {10.1038/342773a0}, \href
  {http://adsabs.harvard.edu/abs/1989Natur.342..773M} {342, 773}

\bibitem[\protect\citeauthoryear{{Molteni}, {Sponholz}  \&
  {Chakrabarti}}{{Molteni} et~al.}{1996}]{molteni96}
{Molteni} D.,  {Sponholz} H.,   {Chakrabarti} S.~K.,  1996, \mn@doi [\apj]
  {10.1086/176775}, \href {http://adsabs.harvard.edu/abs/1996ApJ...457..805M}
  {457, 805}

\bibitem[\protect\citeauthoryear{{Mu{\~n}oz-Darias}, {Motta}, {Stiele}  \&
  {Belloni}}{{Mu{\~n}oz-Darias} et~al.}{2011}]{2011MNRAS.415..292M}
{Mu{\~n}oz-Darias} T.,  {Motta} S.,  {Stiele} H.,   {Belloni} T.~M.,  2011,
  \mn@doi [\mnras] {10.1111/j.1365-2966.2011.18702.x}, \href
  {http://adsabs.harvard.edu/abs/2011MNRAS.415..292M} {415, 292}

\bibitem[\protect\citeauthoryear{{Mushtukov}, {Ingram}  \& {van der
  Klis}}{{Mushtukov} et~al.}{2018}]{2018MNRAS.474.2259M}
{Mushtukov} A.~A.,  {Ingram} A.,   {van der Klis} M.,  2018, \mn@doi [\mnras]
  {10.1093/mnras/stx2872}, \href
  {http://adsabs.harvard.edu/abs/2018MNRAS.474.2259M} {474, 2259}

\bibitem[\protect\citeauthoryear{{Narayan} \& {Yi}}{{Narayan} \&
  {Yi}}{1995}]{1995ApJ...452..710N}
{Narayan} R.,  {Yi} I.,  1995, \mn@doi [\apj] {10.1086/176343}, \href
  {http://adsabs.harvard.edu/abs/1995ApJ...452..710N} {452, 710}

\bibitem[\protect\citeauthoryear{{Nowak}, {Wilms}, {Vaughan}, {Dove}  \&
  {Begelman}}{{Nowak} et~al.}{1999a}]{1999ApJ...515..726N}
{Nowak} M.~A.,  {Wilms} J.,  {Vaughan} B.~A.,  {Dove} J.~B.,   {Begelman}
  M.~C.,  1999a, \mn@doi [\apj] {10.1086/307039}, \href
  {http://adsabs.harvard.edu/abs/1999ApJ...515..726N} {515, 726}

\bibitem[\protect\citeauthoryear{{Nowak}, {Wilms}  \& {Dove}}{{Nowak}
  et~al.}{1999b}]{1999ApJ...517..355N}
{Nowak} M.~A.,  {Wilms} J.,   {Dove} J.~B.,  1999b, \mn@doi [\apj]
  {10.1086/307189}, \href {http://adsabs.harvard.edu/abs/1999ApJ...517..355N}
  {517, 355}

\bibitem[\protect\citeauthoryear{{Parker} et~al.,}{{Parker}
  et~al.}{2015}]{parker15}
{Parker} M.~L.,  et~al., 2015, \mn@doi [\apj] {10.1088/0004-637X/808/1/9},
  \href {http://adsabs.harvard.edu/abs/2015ApJ...808....9P} {808, 9}

\bibitem[\protect\citeauthoryear{Paul et~al.,}{Paul et~al.}{1991}]{paul91}
Paul J.,  et~al., 1991, Adv. Space Res., 11, 289

\bibitem[\protect\citeauthoryear{{Paul}, {Bouchet}, {Churazov}  \&
  {Sunyaev}}{{Paul} et~al.}{1996}]{paul96}
{Paul} J.,  {Bouchet} L.,  {Churazov} E.,   {Sunyaev} R.,  1996, \iaucirc,
  \href {http://adsabs.harvard.edu/abs/1996IAUC.6348....1P} {6348}

\bibitem[\protect\citeauthoryear{{Poutanen}}{{Poutanen}}{2002}]{2002MNRAS.332..257P}
{Poutanen} J.,  2002, \mn@doi [\mnras] {10.1046/j.1365-8711.2002.05272.x},
  \href {http://adsabs.harvard.edu/abs/2002MNRAS.332..257P} {332, 257}

\bibitem[\protect\citeauthoryear{{Poutanen} \& {Coppi}}{{Poutanen} \&
  {Coppi}}{1998}]{1998PhST...77...57P}
{Poutanen} J.,  {Coppi} P.~S.,  1998, Physica Scripta T, \href
  {http://adsabs.harvard.edu/abs/1998PhST...77...57P} {77, 57}

\bibitem[\protect\citeauthoryear{{Poutanen} \& {Fabian}}{{Poutanen} \&
  {Fabian}}{1999}]{1999MNRAS.306L..31P}
{Poutanen} J.,  {Fabian} A.~C.,  1999, \mn@doi [\mnras]
  {10.1046/j.1365-8711.1999.02735.x}, \href
  {http://adsabs.harvard.edu/abs/1999MNRAS.306L..31P} {306, L31}

\bibitem[\protect\citeauthoryear{{Poutanen} \& {Veledina}}{{Poutanen} \&
  {Veledina}}{2014}]{2014SSRv..183...61P}
{Poutanen} J.,  {Veledina} A.,  2014, \mn@doi [\ssr]
  {10.1007/s11214-013-0033-3}, \href
  {http://adsabs.harvard.edu/abs/2014SSRv..183...61P} {183, 61}

\bibitem[\protect\citeauthoryear{{Poutanen} \& {Vurm}}{{Poutanen} \&
  {Vurm}}{2009}]{PV09}
{Poutanen} J.,  {Vurm} I.,  2009, \mn@doi [\apjl]
  {10.1088/0004-637X/690/2/L97}, \href
  {http://adsabs.harvard.edu/abs/2009ApJ...690L..97P} {690, L97}

\bibitem[\protect\citeauthoryear{{Poutanen}, {Krolik}  \& {Ryde}}{{Poutanen}
  et~al.}{1997}]{PKR97}
{Poutanen} J.,  {Krolik} J.~H.,   {Ryde} F.,  1997, \mnras, \href
  {http://adsabs.harvard.edu/abs/1997MNRAS.292L..21P} {292, L21}

\bibitem[\protect\citeauthoryear{{Poutanen}, {Veledina}  \&
  {Zdziarski}}{{Poutanen} et~al.}{2018}]{PVZ18}
{Poutanen} J.,  {Veledina} A.,   {Zdziarski} A.~A.,  2018, \mn@doi [\aap]
  {10.1051/0004-6361/201732345}, \href
  {http://adsabs.harvard.edu/abs/2017arXiv171108509P} {614, A79}

\bibitem[\protect\citeauthoryear{{Reig}, {Mart{\'{\i}}nez-N{\'u}{\~n}ez}  \&
  {Reglero}}{{Reig} et~al.}{2006}]{2006A&A...449..703R}
{Reig} P.,  {Mart{\'{\i}}nez-N{\'u}{\~n}ez} S.,   {Reglero} V.,  2006, \mn@doi
  [\aap] {10.1051/0004-6361:20054289}, \href
  {http://adsabs.harvard.edu/abs/2006A%26A...449..703R} {449, 703}

\bibitem[\protect\citeauthoryear{{Reig}, {Kylafis}, {Papadakis}  \&
  {Costado}}{{Reig} et~al.}{2018}]{2018MNRAS.473.4644R}
{Reig} P.,  {Kylafis} N.~D.,  {Papadakis} I.~E.,   {Costado} M.~T.,  2018,
  \mn@doi [\mnras] {10.1093/mnras/stx2683}, \href
  {http://adsabs.harvard.edu/abs/2018MNRAS.473.4644R} {473, 4644}

\bibitem[\protect\citeauthoryear{{Remillard} \& {McClintock}}{{Remillard} \&
  {McClintock}}{2006}]{remillard06}
{Remillard} R.~A.,  {McClintock} J.~E.,  2006, \mn@doi [\araa]
  {10.1146/annurev.astro.44.051905.092532}, \href
  {http://adsabs.harvard.edu/abs/2006ARA%26A..44...49R} {44, 49}

\bibitem[\protect\citeauthoryear{{Remillard}, {Sobczak}, {Muno}  \&
  {McClintock}}{{Remillard} et~al.}{2002}]{Remillard02}
{Remillard} R.~A.,  {Sobczak} G.~J.,  {Muno} M.~P.,   {McClintock} J.~E.,
  2002, \mn@doi [\apj] {10.1086/324276}, \href
  {http://adsabs.harvard.edu/abs/2002ApJ...564..962R} {564, 962}

\bibitem[\protect\citeauthoryear{{Revnivtsev}, {Gilfanov}  \&
  {Churazov}}{{Revnivtsev} et~al.}{1999}]{RGC99}
{Revnivtsev} M.,  {Gilfanov} M.,   {Churazov} E.,  1999, \aap, \href
  {http://adsabs.harvard.edu/abs/1999A%26A...347L..23R} {347, L23}

\bibitem[\protect\citeauthoryear{{Revnivtsev}, {Gilfanov}  \&
  {Churazov}}{{Revnivtsev} et~al.}{2001}]{Revnivtsev01}
{Revnivtsev} M.,  {Gilfanov} M.,   {Churazov} E.,  2001, \mn@doi [\aap]
  {10.1051/0004-6361:20011413}, \href
  {http://adsabs.harvard.edu/abs/2001A%26A...380..520R} {380, 520}

\bibitem[\protect\citeauthoryear{{Revnivtsev}, {Molkov}  \&
  {Pavlinsky}}{{Revnivtsev} et~al.}{2015}]{2015MNRAS.451.4253R}
{Revnivtsev} M.~G.,  {Molkov} S.~V.,   {Pavlinsky} M.~N.,  2015, \mn@doi
  [\mnras] {10.1093/mnras/stv1263}, \href
  {http://adsabs.harvard.edu/abs/2015MNRAS.451.4253R} {451, 4253}

\bibitem[\protect\citeauthoryear{{Ross} \& {Fabian}}{{Ross} \&
  {Fabian}}{2005}]{ross05}
{Ross} R.~R.,  {Fabian} A.~C.,  2005, \mn@doi [\mnras]
  {10.1111/j.1365-2966.2005.08797.x}, \href
  {http://adsabs.harvard.edu/abs/2005MNRAS.358..211R} {358, 211}

\bibitem[\protect\citeauthoryear{{Schultheis} et~al.,}{{Schultheis}
  et~al.}{2014}]{2014A&A...566A.120S}
{Schultheis} M.,  et~al., 2014, \mn@doi [\aap] {10.1051/0004-6361/201322788},
  \href {http://adsabs.harvard.edu/abs/2014A%26A...566A.120S} {566, A120}

\bibitem[\protect\citeauthoryear{{Shakura} \& {Sunyaev}}{{Shakura} \&
  {Sunyaev}}{1973}]{shakura73}
{Shakura} N.~I.,  {Sunyaev} R.~A.,  1973, \aap, \href
  {http://adsabs.harvard.edu/abs/1973A%26A....24..337S} {24, 337}

\bibitem[\protect\citeauthoryear{{Shapiro}, {Lightman}  \& {Eardley}}{{Shapiro}
  et~al.}{1976}]{1976ApJ...204..187S}
{Shapiro} S.~L.,  {Lightman} A.~P.,   {Eardley} D.~M.,  1976, \mn@doi [\apj]
  {10.1086/154162}, \href {http://adsabs.harvard.edu/abs/1976ApJ...204..187S}
  {204, 187}

\bibitem[\protect\citeauthoryear{{Shidatsu} et~al.,}{{Shidatsu}
  et~al.}{2014}]{2014ApJ...789..100S}
{Shidatsu} M.,  et~al., 2014, \mn@doi [\apj] {10.1088/0004-637X/789/2/100},
  \href {http://adsabs.harvard.edu/abs/2014ApJ...789..100S} {789, 100}

\bibitem[\protect\citeauthoryear{{Sobczak}, {McClintock}, {Remillard}, {Cui},
  {Levine}, {Morgan}, {Orosz}  \& {Bailyn}}{{Sobczak}
  et~al.}{2000}]{2000ApJ...531..537S}
{Sobczak} G.~J.,  {McClintock} J.~E.,  {Remillard} R.~A.,  {Cui} W.,  {Levine}
  A.~M.,  {Morgan} E.~H.,  {Orosz} J.~A.,   {Bailyn} C.~D.,  2000, \mn@doi
  [\apj] {10.1086/308463}, \href
  {http://adsabs.harvard.edu/abs/2000ApJ...531..537S} {531, 537}

\bibitem[\protect\citeauthoryear{{Tagger} \& {Pellat}}{{Tagger} \&
  {Pellat}}{1999}]{tagger99}
{Tagger} M.,  {Pellat} R.,  1999, \aap, \href
  {http://adsabs.harvard.edu/abs/1999A%26A...349.1003T} {349, 1003}

\bibitem[\protect\citeauthoryear{{Tanaka} \& {Shibazaki}}{{Tanaka} \&
  {Shibazaki}}{1996}]{tanaka96}
{Tanaka} Y.,  {Shibazaki} N.,  1996, \mn@doi [\araa]
  {10.1146/annurev.astro.34.1.607}, \href
  {http://adsabs.harvard.edu/abs/1996ARA%26A..34..607T} {34, 607}

\bibitem[\protect\citeauthoryear{{Terrell}}{{Terrell}}{1972}]{1972ApJ...174L..35T}
{Terrell} Jr. N.~J.,  1972, \mn@doi [\apjl] {10.1086/180944}, \href
  {http://adsabs.harvard.edu/abs/1972ApJ...174L..35T} {174, L35}

\bibitem[\protect\citeauthoryear{{Titarchuk} \& {Fiorito}}{{Titarchuk} \&
  {Fiorito}}{2004}]{2004ApJ...612..988T}
{Titarchuk} L.,  {Fiorito} R.,  2004, \mn@doi [\apj] {10.1086/422573}, \href
  {http://adsabs.harvard.edu/abs/2004ApJ...612..988T} {612, 988}

\bibitem[\protect\citeauthoryear{{Uttley}, {Cackett}, {Fabian}, {Kara}  \&
  {Wilkins}}{{Uttley} et~al.}{2014}]{2014A&ARv..22...72U}
{Uttley} P.,  {Cackett} E.~M.,  {Fabian} A.~C.,  {Kara} E.,   {Wilkins} D.~R.,
  2014, \mn@doi [\aapr] {10.1007/s00159-014-0072-0}, \href
  {http://adsabs.harvard.edu/abs/2014A%26ARv..22...72U} {22, 72}

\bibitem[\protect\citeauthoryear{{Vaughan} \& {Nowak}}{{Vaughan} \&
  {Nowak}}{1997}]{1997ApJ...474L..43V}
{Vaughan} B.~A.,  {Nowak} M.~A.,  1997, \mn@doi [\apjl] {10.1086/310430}, \href
  {http://adsabs.harvard.edu/abs/1997ApJ...474L..43V} {474, L43}

\bibitem[\protect\citeauthoryear{{Veledina}, {Poutanen}  \& {Vurm}}{{Veledina}
  et~al.}{2011}]{VPV11}
{Veledina} A.,  {Poutanen} J.,   {Vurm} I.,  2011, \mn@doi [\apjl]
  {10.1088/2041-8205/737/1/L17}, \href
  {http://adsabs.harvard.edu/abs/2011ApJ...737L..17V} {737, L17}

\bibitem[\protect\citeauthoryear{{Veledina}, {Poutanen}  \& {Vurm}}{{Veledina}
  et~al.}{2013a}]{VPV13}
{Veledina} A.,  {Poutanen} J.,   {Vurm} I.,  2013a, \mn@doi [\mnras]
  {10.1093/mnras/stt124}, \href
  {http://adsabs.harvard.edu/abs/2013MNRAS.430.3196V} {430, 3196}

\bibitem[\protect\citeauthoryear{{Veledina}, {Poutanen}  \&
  {Ingram}}{{Veledina} et~al.}{2013b}]{VPI13}
{Veledina} A.,  {Poutanen} J.,   {Ingram} A.,  2013b, \mn@doi [\apj]
  {10.1088/0004-637X/778/2/165}, \href
  {http://adsabs.harvard.edu/abs/2013ApJ...778..165V} {778, 165}

\bibitem[\protect\citeauthoryear{{Veledina}, {Revnivtsev}, {Durant}, {Gandhi}
  \& {Poutanen}}{{Veledina} et~al.}{2015}]{VRD15}
{Veledina} A.,  {Revnivtsev} M.~G.,  {Durant} M.,  {Gandhi} P.,   {Poutanen}
  J.,  2015, \mn@doi [\mnras] {10.1093/mnras/stv2201}, \href
  {http://adsabs.harvard.edu/abs/2015MNRAS.454.2855V} {454, 2855}

\bibitem[\protect\citeauthoryear{{Verner}, {Ferland}, {Korista}  \&
  {Yakovlev}}{{Verner} et~al.}{1996}]{verner96}
{Verner} D.~A.,  {Ferland} G.~J.,  {Korista} K.~T.,   {Yakovlev} D.~G.,  1996,
  \mn@doi [\apj] {10.1086/177435}, \href
  {http://adsabs.harvard.edu/abs/1996ApJ...465..487V} {465, 487}

\bibitem[\protect\citeauthoryear{{Vignarca}, {Migliari}, {Belloni}, {Psaltis}
  \& {van der Klis}}{{Vignarca} et~al.}{2003}]{vignarca03}
{Vignarca} F.,  {Migliari} S.,  {Belloni} T.,  {Psaltis} D.,   {van der Klis}
  M.,  2003, \mn@doi [\aap] {10.1051/0004-6361:20021542}, \href
  {http://adsabs.harvard.edu/abs/2003A%26A...397..729V} {397, 729}

\bibitem[\protect\citeauthoryear{{Vikhlinin}, {Churazov}  \&
  {Gilfanov}}{{Vikhlinin} et~al.}{1994}]{1994A&A...287...73V}
{Vikhlinin} A.,  {Churazov} E.,   {Gilfanov} M.,  1994, \aap, \href
  {http://adsabs.harvard.edu/abs/1994A%26A...287...73V} {287, 73}

\bibitem[\protect\citeauthoryear{{Wijnands} \& {van der Klis}}{{Wijnands} \&
  {van der Klis}}{1999}]{wijnands99}
{Wijnands} R.,  {van der Klis} M.,  1999, \mn@doi [\apj] {10.1086/306993},
  \href {http://adsabs.harvard.edu/abs/1999ApJ...514..939W} {514, 939}

\bibitem[\protect\citeauthoryear{{Wijnands}, {M{\'e}ndez}, {Miller}  \&
  {Homan}}{{Wijnands} et~al.}{2001}]{2001MNRAS.328..451W}
{Wijnands} R.,  {M{\'e}ndez} M.,  {Miller} J.~M.,   {Homan} J.,  2001, \mn@doi
  [\mnras] {10.1046/j.1365-8711.2001.04871.x}, \href
  {http://adsabs.harvard.edu/abs/2001MNRAS.328..451W} {328, 451}

\bibitem[\protect\citeauthoryear{{Wilms}, {Allen}  \& {McCray}}{{Wilms}
  et~al.}{2000}]{wilms00}
{Wilms} J.,  {Allen} A.,   {McCray} R.,  2000, \mn@doi [\apj] {10.1086/317016},
  \href {http://adsabs.harvard.edu/abs/2000ApJ...542..914W} {542, 914}

\bibitem[\protect\citeauthoryear{{Yan} \& {Yu}}{{Yan} \& {Yu}}{2017}]{yan17}
{Yan} Z.,  {Yu} W.,  2017, \mn@doi [\mnras] {10.1093/mnras/stx1562}, \href
  {http://adsabs.harvard.edu/abs/2017MNRAS.470.4298Y} {470, 4298}

\bibitem[\protect\citeauthoryear{{Yan} et~al.,}{{Yan}
  et~al.}{2017}]{2017MNRAS.465.1926Y}
{Yan} S.-P.,  et~al., 2017, \mn@doi [\mnras] {10.1093/mnras/stw2916}, \href
  {http://adsabs.harvard.edu/abs/2017MNRAS.465.1926Y} {465, 1926}

\bibitem[\protect\citeauthoryear{{Zdziarski} \& {Gierli{\'n}ski}}{{Zdziarski}
  \& {Gierli{\'n}ski}}{2004}]{ZG04}
{Zdziarski} A.~A.,  {Gierli{\'n}ski} M.,  2004, \mn@doi [Progr. Theor. Phys.
  Suppl.] {10.1143/PTPS.155.99}, \href
  {http://adsabs.harvard.edu/abs/2004PThPS.155...99Z} {155, 99}

\bibitem[\protect\citeauthoryear{{Zdziarski}, {Grove}, {Poutanen}, {Rao}  \&
  {Vadawale}}{{Zdziarski} et~al.}{2001}]{ZGP01}
{Zdziarski} A.~A.,  {Grove} J.~E.,  {Poutanen} J.,  {Rao} A.~R.,   {Vadawale}
  S.~V.,  2001, \mn@doi [\apjl] {10.1086/320932}, \href
  {http://adsabs.harvard.edu/abs/2001ApJ...554L..45Z} {554, L45}

\bibitem[\protect\citeauthoryear{{Zhang}, {Jahoda}, {Swank}, {Morgan}  \&
  {Giles}}{{Zhang} et~al.}{1995}]{1995ApJ...449..930Z}
{Zhang} W.,  {Jahoda} K.,  {Swank} J.~H.,  {Morgan} E.~H.,   {Giles} A.~B.,
  1995, \mn@doi [\apj] {10.1086/176111}, \href
  {http://adsabs.harvard.edu/abs/1995ApJ...449..930Z} {449, 930}

\bibitem[\protect\citeauthoryear{{Zhang}, {Wang}, {M{\'e}ndez}, {Chen}, {Qu},
  {Altamirano}  \& {Belloni}}{{Zhang} et~al.}{2017}]{2017ApJ...845..143Z}
{Zhang} L.,  {Wang} Y.,  {M{\'e}ndez} M.,  {Chen} L.,  {Qu} J.,  {Altamirano}
  D.,   {Belloni} T.,  2017, \mn@doi [\apj] {10.3847/1538-4357/aa8138}, \href
  {http://adsabs.harvard.edu/abs/2017ApJ...845..143Z} {845, 143}

\bibitem[\protect\citeauthoryear{{van den Eijnden}, {Ingram}, {Uttley},
  {Motta}, {Belloni}  \& {Gardenier}}{{van den Eijnden}
  et~al.}{2017}]{2017MNRAS.464.2643V}
{van den Eijnden} J.,  {Ingram} A.,  {Uttley} P.,  {Motta} S.~E.,  {Belloni}
  T.~M.,   {Gardenier} D.~W.,  2017, \mn@doi [\mnras] {10.1093/mnras/stw2634},
  \href {http://adsabs.harvard.edu/abs/2017MNRAS.464.2643V} {464, 2643}

\makeatother
\end{thebibliography}
\bsp	
\label{lastpage}
\end{document}